\documentclass[lettersize,journal]{IEEEtran}
\usepackage{amsmath,amsfonts,amssymb}
\usepackage{algorithm}
\usepackage{algpseudocode}
\usepackage{array}
\usepackage{subfigure}
\usepackage[caption=false,font=normalsize,labelfont=sf,textfont=sf]{subfig}
\usepackage{textcomp}
\usepackage{stfloats}
\usepackage{url}
\usepackage{verbatim}
\usepackage{graphicx}
\usepackage{cite}
\usepackage{amsthm}

\newtheorem{lemma}{Lemma}

\begin{document}

\title{Location-Aware Performance Analysis of HAPS-RIS-Assisted Multiple-Antenna Systems with Practical RIS Impairments}

\author{Tayfun Yilmaz, Haci Ilhan, Ali Gorcin and Halim Yanikomeroglu
\thanks{}
}

\markboth{Possible IEEE Journal,~Vol.~XX, No.~X, Month~2026}%
{Shell \MakeLowercase{\textit{et al.}}: A Sample Article Using IEEEtran.cls for IEEE Journals}


\maketitle

\begin{abstract}

High-altitude platform station (HAPS)-mounted reconfigurable intelligent surfaces (RISs) offer a promising means of extending the coverage and reliability of future 6G and beyond wireless networks, but their performance is strongly affected by cascaded line-of-sight (LoS)-dominant fading, practical RIS hardware impairments, and location-dependent atmospheric attenuation. This paper develops a tractable performance-analysis framework for HAPS-RIS-assisted multiple-antenna systems over cascaded Rician--Rician channels. The small-scale effective channel power is statistically characterized by exploiting the structured covariance of the cascaded channel and applying saddlepoint approximation (SPA), while the composite log-normal large-scale fading and atmosphere-dependent attenuation are incorporated through Gauss–Hermite quadrature to obtain tractable approximations for the received-SNR distribution and outage probability. The framework further accounts for finite-resolution RIS phase shifts and phase-dependent reflection amplitudes. To connect the analysis with a realistic deployment, the transmitter, receiver, and HAPS locations are mapped through Earth-centered Earth-fixed/east-north-up (ECEF/ENU) coordinate transformations to obtain the propagation directions and array steering vectors, enabling geometry-aided transmit precoding and LoS-aligned RIS phase configuration without requiring full instantaneous channel state information (CSI). Monte Carlo simulations show excellent agreement with the proposed analytical probability density function (PDF) and outage expressions. Moreover, the geometry-aided LoS-aligned precoder achieves outage performance nearly indistinguishable from instantaneous dominant-eigenmode precoding. The results further reveal that coherent RIS phase alignment is essential for reducing the required transmit power, larger RIS apertures are needed under severe atmospheric attenuation, and increasing the RIS phase resolution beyond a few bits provides only limited additional gain.
\end{abstract}
\IEEEpubidadjcol

\begin{IEEEkeywords}
HAPS-mounted RIS, multiple-antenna systems, geometry-aided beamforming, outage probability, cascaded Rician--Rician fading, atmospheric attenuation, practical RIS impairments, saddlepoint approximation.
\end{IEEEkeywords}

\section{INTRODUCTION}
\label{Sec_Intro}
\IEEEpubidadjcol 

The vision for future 6G and beyond wireless networks is to provide seamless connectivity for everyone, everything, and everywhere. In this context, aerial platforms such as high-altitude platform stations (HAPS) have emerged as promising solutions to extend wireless coverage and enable connectivity in hard-to-reach or underserved regions~\cite{G_Kurt_HAPS, HAPS_Connectivity}. Recently, HAPS-mounted reconfigurable intelligent surfaces (RIS) architectures, referred to as HAPS-RIS, have
attracted increasing attention as a promising solution for coverage extension and link reliability enhancement, since they can provide controllable reflection paths and reconfigure the wireless propagation environment in otherwise challenging deployment scenarios \cite{HAPS_Intro1,HAPS_Intro2, HAPS_RIS_Magazine_1,HAPS_RIS_Intro, HAPS_RIS_Magazine_2}. Despite their promising potential, the performance of HAPS-RIS systems strongly depends on the underlying channel statistics, RIS configuration, deployment geometry, and transceiver design. Therefore, a rigorous
performance characterization is necessary to identify their practical design limits. In this direction, several recent studies have investigated HAPS-RIS systems from different perspectives, including channel modeling, coverage extension, beamforming optimization, and reliability analysis \cite{3,1,2,HAPS_RIS_NOMA,6}.

\subsection{Related Works and Motivation}

Early studies established the link-budget advantage of hosting large reflecting surfaces on HAPS~\cite{AHO-13} and investigated HAPS-RIS resource allocation, beyond-cell coverage, and aerodynamic-HAPS phase design~\cite{HAPS_RIS_Intro,HAPS_Intro2,AHO-14}. Performance analyses of HAPS-mounted surfaces remain scarce. The multi-user network of \cite{3} and the secrecy analysis of \cite{1} model the RIS with unit-amplitude reflection and ideal phase shifts and assume single-antenna terminals. The outage probability (OP) of a HAPS-RIS-assisted UAV link with a residual direct path is analyzed in \cite{2}, with emphasis on inter-symbol interference; \cite{AHO-15} analyzes an integrated satellite--HAPS--ground link with a HAPS-mounted RIS using a central-limit approximation; \cite{AHO-16,AHO-17} characterize RIS-assisted hybrid satellite--terrestrial links and dual aerial-RIS GEO links through generalized-\(K\) fits, while assuming single-antenna transmitter and receiver terminals and unit-amplitude RIS elements; and \cite{AHO-18} treats a THz HAPS-RIS link with a bounded pointing-error factor. Multiple-access and optimization-oriented studies (active-RIS beamspace NOMA~\cite{HAPS_RIS_NOMA}, NOMA-RIS~\cite{6}, secure RSMA~\cite{AHO-19}) rely on instantaneous channel state information (CSI) and single-antenna ground terminals. None of these works includes standardized Ka-band atmospheric attenuation or practical RIS impairment effects, and they target spectral efficiency or power allocation rather than the received-SNR statistics.
\IEEEpubidadjcol

Since HAPS-RIS architectures can be viewed as aerial realizations of RIS-assisted multiple-antenna systems, the broader analytical literature on RIS-aided multiple-input multiple-output (MIMO) communications provides useful insights. When the cascaded channel follows Rayleigh fading, the resulting channel statistics are more tractable, allowing conventional random-matrix-based tools to be applied for performance analysis \cite{abbasi2024ergodic, TVT_Tracy_Widom}. However, aerial-to-ground links generally exhibit LoS-dominant propagation and are commonly modeled using Rician fading. Therefore, studies considering cascaded Rician-Rician fading channels are particularly relevant to the performance characterization of HAPS-RIS systems. In \cite{4}, the ergodic capacity (EC) of RIS-assisted multiple-antenna systems is analyzed under a Rayleigh--Rician cascaded channel assuming unit-amplitude RIS reflection and uniform phase shifts, with the analysis mainly focusing on relatively small RIS sizes. The capacity maximization problem under Rician fading is investigated in \cite{5}; however, the users are equipped with single antennas and the RIS is again modeled with unit-amplitude reflection. More recently, \cite{Mosleh_2026} derived tight approximations and bounds for the EC of RIS-aided multiple-antenna systems with amplitude--phase coupling, but considered a Rayleigh--Rician cascaded channel and an eigenvalue-based capacity formulation rather than a direct statistical characterization of the received SNR.

Therefore, an SNR-statistics-based analytical framework for HAPS-RIS-assisted multiple-antenna systems that jointly accounts for arbitrary transmit/receive antenna configurations, cascaded Rician--Rician fading, log-normal shadowing, practical RIS responses, and location-dependent atmospheric propagation effects is still lacking in the literature. In particular, the received SNR statistics become highly non-trivial due to the joint impact of cascaded Rician small-scale fading, strong deterministic LoS components, aerial-to-ground shadowing, practical RIS phase/amplitude responses, and atmospheric attenuation. Moreover, unlike conventional terrestrial RIS deployments, the HAPS link budget is strongly influenced by the geographical deployment, elevation-dependent propagation conditions, and long-term atmospheric effects, which should be incorporated into the reliability analysis.

From a statistical perspective, the transmitter--RIS and RIS--receiver links are coupled through the LoS-dominant structure and the common RIS response, so the resulting cascaded channel does not generally admit the independence structure required by conventional non-central Wishart-based formulations. Meanwhile, the deterministic LoS geometry provides useful directional information that can be exploited for low-complexity transmission without relying on full instantaneous CSI. These characteristics motivate an outage-oriented analysis that jointly accounts for geometry-aided LoS transmission, cascaded small-scale fading, practical RIS responses, and location-dependent large-scale propagation effects in HAPS-RIS-assisted multiple-antenna systems.

\subsection{Contributions and Paper Organization}

Motivated by the aforementioned limitations, this paper develops a tractable analytical framework for the outage characterization of HAPS-RIS-assisted multiple-antenna systems under practical propagation and RIS hardware conditions. The framework integrates geometry-aware transmission with statistical SNR analysis and enables location-specific reliability evaluation for HAPS-RIS deployments.

The main contributions of this paper are summarized as follows:

\begin{itemize}

\item We formulate a geometry-aware HAPS-RIS-assisted multiple-antenna system model in which the geographical coordinates of the transmitter, receiver, and HAPS are transformed through Earth-centered Earth-fixed/east-north-up (ECEF/ENU) coordinates to determine the corresponding propagation directions and array steering vectors. These geometry-derived quantities are then exploited to construct a LoS-aligned transmit precoder and a coherent RIS phase configuration that strengthens the deterministic LoS component without requiring full instantaneous CSI. The model further incorporates cascaded Rician--Rician fading, log-normal shadowing, and practical phase-dependent RIS amplitude responses.

\item We derive a tractable statistical characterization of the effective cascaded channel under the proposed LoS-aligned transmission strategy. By exploiting the structured covariance induced by the Rician--Rician channel and the common RIS response, the effective channel power is decomposed into one noncentral dominant mode and \(N_r-1\) central scattering modes. This structure enables a compact moment generating function (MGF) and cumulant generating function (CGF) representation, from which saddlepoint-based approximations (SPA) for the probability density function (PDF) and cumulative distribution function (CDF) of the effective small-scale channel power are obtained.

\item We incorporate the large-scale HAPS propagation effects into the received-SNR statistics by jointly accounting for log-normal shadowing and location-dependent atmospheric attenuation, including gaseous absorption, rain, cloud, and tropospheric scintillation losses. Using the resulting log-normal representation of the composite large-scale gain, Gauss--Hermite quadrature is employed to combine these effects with the SPA-based small-scale characterization, yielding tractable analytical expressions for the received-SNR distribution and outage probability under prescribed atmospheric exceedance conditions.

\item We conduct extensive numerical evaluations and Monte Carlo simulations to validate the proposed analytical framework and to extract practical design insights. The derived PDF and outage expressions closely match the simulation results over the considered RIS configurations and atmospheric conditions. Moreover, the geometry-aided LoS-aligned precoder achieves outage performance nearly identical to instantaneous dominant-eigenmode precoding, while requiring substantially less CSI and computational complexity. The results further quantify the joint impact of RIS aperture, coherent phase alignment, phase resolution, transmit power, and atmospheric severity, showing that large RIS deployments can effectively compensate for severe propagation losses only when the reflected components are coherently combined. In addition, the gain from increasing the RIS phase resolution becomes marginal beyond a few bits.

\end{itemize}

The remainder of this paper is organized as follows. Section~\ref{Sec:sys_model} introduces the considered HAPS-RIS-assisted multiple-antenna system model, including the communication link structure, large- and small-scale channel models, RIS configuration, and practical RIS response parameters. Section~\ref{Sec:performance_analysis} presents the statistical characterization of the effective cascaded channel, formulates the received SNR model, and develops the SPA-based PDF, CDF, and outage probability expressions together with the Gauss–Hermite quadratures. Section~\ref{Sec:simulation_results} describes the considered geographical deployment, atmospheric attenuation calculations, simulation parameters, and numerical results, and provides practical design insights. Finally, Section~\ref{Sec:conclusions} concludes the paper.
\begin{figure}[t!]
    \centering
    \includegraphics[width=\columnwidth]{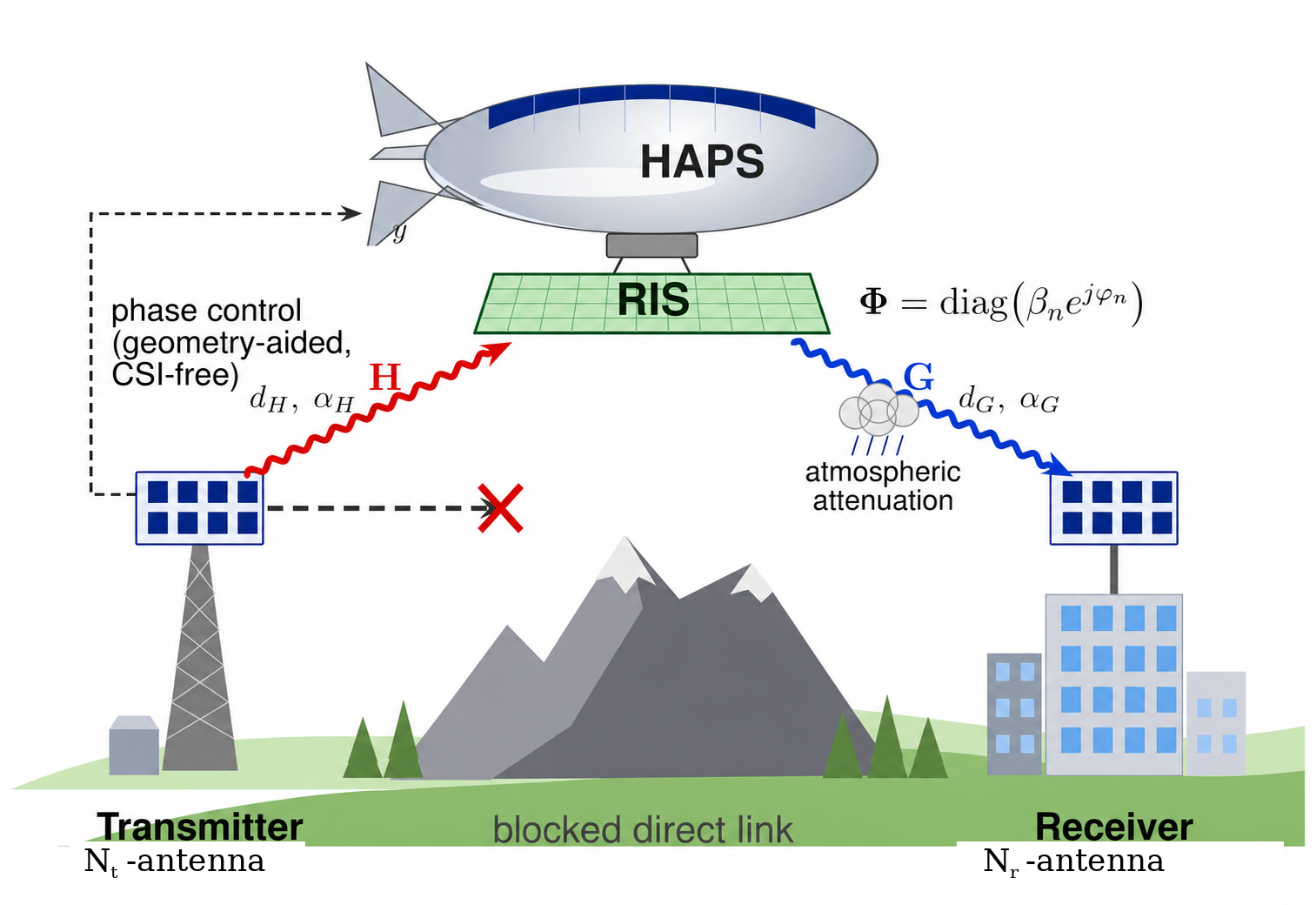}
    \caption{Considered HAPS-mounted RIS-assisted multiple-antenna system.}
    \label{fig:HAPS_system}
\end{figure}

\subsection{Notation}

Upper-case and lower-case boldface letters denote matrices and vectors, respectively. The operators \((\cdot)^T\), \((\cdot)^H\), \((\cdot)^*\), and \(\otimes\) denote transpose, Hermitian transpose, complex conjugate, and Kronecker product, respectively. The notations \(\mathbb{E}[\cdot]\), \(\mathrm{Var}(\cdot)\), \(\|\cdot\|\), \(|\cdot|\), \(\mathrm{tr}(\cdot)\), \(\mathrm{diag}(\cdot)\), and \(\angle(\cdot)\) denote expectation, variance, Euclidean norm, scalar magnitude, trace, diagonal matrix construction, and complex phase, respectively. Moreover, \(\mathcal{CN}(\boldsymbol{\mu},\mathbf{\Sigma})\) denotes a circularly symmetric complex Gaussian random vector with mean \(\boldsymbol{\mu}\) and covariance \(\mathbf{\Sigma}\), \(\mathcal{N}(\mu,\sigma^2)\) denotes a real Gaussian random variable, \(\mathbf{I}_N\) denotes the \(N\times N\) identity matrix, and \(j=\sqrt{-1}\). The functions \(\ln(\cdot)\), \(\log_{10}(\cdot)\), and \(\exp(\cdot)\) denote the natural logarithm, base-10 logarithm, and exponential function, respectively, while \(\operatorname{sign}(\cdot)\), \(\arg\max\), \(\arg\min\), and \(\triangleq\) denote the sign function, maximization argument, minimization argument, and equality by definition, respectively. Finally, \(\Psi(\cdot)\) and \(\psi(\cdot)\) denote the standard normal CDF and PDF, respectively.

\section{System Model}
\label{Sec:sys_model}

As illustrated in Fig.~\ref{fig:HAPS_system}, we consider a HAPS-mounted
RIS-assisted single-stream multiple-antenna system, where a terrestrial transmitter
(T) equipped with $N_t$ antennas communicates with a terrestrial receiver
(R) equipped with $N_r$ antennas through an RIS comprising
$N_{\rm RIS}$ reflecting elements deployed on a HAPS.
The direct T--R link is assumed to be unavailable, and communication is
established through the T--RIS and RIS--R links. Both terrestrial
terminals operate outdoors and are positioned such that unobstructed
LoS propagation toward the HAPS-mounted RIS is maintained. The
transmitter, RIS and the receiver employ uniform rectangular arrays (URAs).

\subsection{Channel Model}

The T--RIS and RIS--R channels are denoted by
$\mathbf{H}\in\mathbb{C}^{N_{\rm RIS}\times N_t}$ and
$\mathbf{G}\in\mathbb{C}^{N_r\times N_{\rm RIS}}$, respectively.
Each channel is decomposed into its large- and small-scale components as
\begin{equation}
\mathbf{H}
=
\sqrt{L_H}\widehat{\mathbf{H}},
\qquad
\mathbf{G}
=
\sqrt{L_G}\widehat{\mathbf{G}},
\label{eq:channel_decomposition}
\end{equation}
where $L_H$ and $L_G$ denote the large-scale channel gains, while
$\widehat{\mathbf{H}}$ and $\widehat{\mathbf{G}}$ represent the
corresponding small-scale fading matrices.

\subsubsection{Large-scale fading model}For a general satellite- or HAPS-to-ground link, 3GPP TR~38.811
expresses the large-scale propagation loss in terms of the basic path
loss, atmospheric propagation losses, and building-entry loss
\cite{3GPP_TR_38811}. For compactness, we write
\begin{equation}
PL_\ell
=
PL_{b,\ell}
+
PL_{A,\ell}
+
PL_{e,\ell},
\qquad
\ell\in\{H,G\},
\label{eq:general_HAPS_PL}
\end{equation}
where $PL_{b,\ell}$ denotes the basic path loss, $PL_{A,\ell}$
collects the atmospheric attenuation terms, and $PL_{e,\ell}$ denotes
the building-entry loss. The basic path loss is given by
\begin{equation}
PL_{b,\ell}
=
FSPL(d_\ell,f_c)
+
\chi_\ell
+
CL_\ell,
\label{eq:basic_PL}
\end{equation}
where
\begin{equation}
FSPL(d_\ell,f_c)
=
32.45
+
20\log_{10}(f_c)
+
20\log_{10}(d_\ell),
\label{eq:FSPL}
\end{equation}
with $f_c$ in GHz and $d_\ell$ in meters. Moreover,
\(
\chi_\ell
\sim
\mathcal{N}(0,\sigma_{\chi,\ell}^2)
\label{eq:shadowing}
\)
denotes the shadow-fading component in the dB domain, while
$CL_\ell$ denotes clutter loss. Since both terrestrial terminals are positioned such that LoS
propagation toward the HAPS-RIS is maintained, the clutter losses are
set to $CL_H=CL_G=0$. Moreover, both terminals are located outdoors,
and therefore $PL_{e,H}=PL_{e,G}=0$. In addition, since the considered HAPS altitude is well below the ionospheric
region, ionospheric scintillation is not included. Accordingly,
$PL_{A,\ell}$ accounts for gaseous attenuation, rain/cloud attenuation,
and tropospheric scintillation. To incorporate their long-term
statistical variations, the atmospheric attenuation is evaluated for a
prescribed exceedance probability $p$ according to ITU-R P.618
\cite{ITU_R_P618}. For link $\ell$, the corresponding attenuation level is
\begin{equation}
PL_{A,\ell}(p)
=
A_{G,\ell}(p)
+
\sqrt{
\left[
A_{R,\ell}(p)+A_{C,\ell}(p)
\right]^2
+
A_{S,\ell}^2(p)
},
\label{eq:atmospheric_loss_lowp}
\end{equation}
for $0.001\%\leq p\leq5\%$, whereas
\begin{equation}
PL_{A,\ell}(p)
=
A_{G,\ell}(p)
+
\sqrt{
A_{C,\ell}^2(p)
+
A_{S,\ell}^2(p)
},
\label{eq:atmospheric_loss_highp}
\end{equation}
for $5\%<p\leq50\%$.
Here, $A_{G,\ell}(p)$, $A_{R,\ell}(p)$,
$A_{C,\ell}(p)$, and $A_{S,\ell}(p)$ denote the gaseous,
rain, cloud, and tropospheric-scintillation attenuation levels,
respectively\footnote{The attenuation components are evaluated using the
corresponding ITU-R models: $A_{R,\ell}(p)$ and $A_{S,\ell}(p)$
according to ITU-R P.618, $A_{C,\ell}(p)$ according to
ITU-R P.840~\cite{ITU_R_P840}, and $A_{G,\ell}(p)$ according to ITU-R P.676~\cite{ITU_R_P676}.
}.

To incorporate the free-space propagation loss in \eqref{eq:basic_PL}
while retaining the RIS-specific scattering and angular responses, the
deterministic path gains of the two hops are expressed as
\cite{abbasi2024ergodic, PL_RIS}
\begin{subequations}
\begin{align}
&
\mathcal{L}_H
=
G_tG_{\rm RIS}F(\theta_H,\phi_H)
10^{-FSPL(d_H,f_c)/10},
\label{eq:Path_Loss_H}
\\
&\mathcal{L}_G
=
G_rG_{\rm RIS}F(\theta_G,\phi_G)
10^{-FSPL(d_G,f_c)/10}.
\label{eq:Path_Loss_G}
\end{align}
\end{subequations}
Using
\(
10^{-FSPL(d,f_c)/10}
=
(\lambda_c/4\pi d)^2,
\)
these expressions are equivalent to the RIS-assisted far-field
path-gain formulation in \cite{abbasi2024ergodic, PL_RIS}. Here, $d_H$ and
$d_G$ denote the corresponding slant ranges, $G_t$ and $G_r$ are the
transmit and receive antenna-element gains, and $G_{\rm RIS}$ denotes the
scattering gain of an RIS element, given by
\(
G_{\rm RIS}
=
{4\pi d_xd_y}/{\lambda_c^2}.
\)
Here, $d_x$ and $d_y$ denote the physical dimensions of an RIS element,
and $\lambda_c$ is the carrier wavelength.
Following~\cite{PL_RIS}, the normalized RIS-element radiation pattern is modeled as
\begin{equation}
F(\theta,\phi)
=
\begin{cases}
\cos^3\theta, & \theta\in\left[0,\frac{\pi}{2}\right],\\
0, & \theta\in\left(\frac{\pi}{2},\pi\right].
\end{cases}
\label{eq:fcos}
\end{equation}
where \(F(\theta_H,\phi_H)\) and \(F(\theta_G,\phi_G)\) represent the responses along the incident and reflected directions, respectively.

Combining the deterministic RIS-assisted path gain with shadow fading
and the atmospheric attenuation evaluated at the prescribed exceedance
probability $p$, the large-scale gain of link
$\ell\in\{H,G\}$ is expressed as
\begin{equation}
L_\ell
=
\mathcal{L}_\ell
10^{-\left[
\chi_\ell+PL_{A,\ell}(p)
\right]/10}.
\label{eq:large_scale_gain}
\end{equation}
Accordingly, the cascaded large-scale gain becomes
\begin{equation}
L_HL_G
=
\mathcal{L}_H\mathcal{L}_G
10^{-\left[
\chi_H+\chi_G
+
PL_{A,H}(p)
+
PL_{A,G}(p)
\right]/10}.
\label{eq:L_total_large_scale}
\end{equation}

\subsubsection{Small-scale fading model}The small-scale fading components of the two hops are modeled as Rician
channels,
\begin{equation}
\widehat{\mathbf{H}}
=
\sqrt{\frac{K_H}{K_H+1}}
\mathbf{H}_{\rm LoS}
+
\sqrt{\frac{1}{K_H+1}}
\mathbf{H}_{\rm NLoS},
\label{eq:rician_H}
\end{equation}
and
\begin{equation}
\widehat{\mathbf{G}}
=
\sqrt{\frac{K_G}{K_G+1}}
\mathbf{G}_{\rm LoS}
+
\sqrt{\frac{1}{K_G+1}}
\mathbf{G}_{\rm NLoS},
\label{eq:rician_G}
\end{equation}
where $K_H$ and $K_G$ denote the Rician $K$-factors of the T--RIS
and RIS--R links, respectively. The scattered components have i.i.d.
entries
\begin{equation}
[\mathbf{H}_{\rm NLoS}]_{n,t}
\sim
\mathcal{CN}(0,\sigma_h^2),
\qquad
[\mathbf{G}_{\rm NLoS}]_{r,n}
\sim
\mathcal{CN}(0,\sigma_g^2).
\end{equation}
The LoS components admit a rank-one structure governed by the array
steering vectors, i.e.,
\begin{subequations}
\begin{align}
&\mathbf H_{\rm LoS}
=
\mathbf a_{\rm RIS}(\phi_H,\theta_H)
\mathbf a_{\rm T}^{H}(\phi_t,\theta_t),
\label{eq:H_LoS}
\\
&
\mathbf G_{\rm LoS}
=
\mathbf a_{\rm R}(\phi_r,\theta_r)
\mathbf a_{\rm RIS}^{H}(\phi_G,\theta_G).
\label{eq:G_LoS}
\end{align}
\end{subequations}
Here, $(\phi_t,\theta_t)$ and $(\phi_H,\theta_H)$ denote the azimuth
and polar AoD and AoA angles of the T--RIS LoS link, respectively,
while $(\phi_G,\theta_G)$ and $(\phi_r,\theta_r)$ denote the
corresponding AoD and AoA angles of the RIS--R LoS link.
For a generic $N_x\times N_y$ URA, the steering vector is expressed as
\begin{equation}
\mathbf a(\phi,\theta)
=
\mathbf a_x(u)\otimes\mathbf a_y(v),
\label{eq:generic_URA}
\end{equation}
where
\(
u=\sin\theta\cos\phi,\,
v=\sin\theta\sin\phi,
\)
with
\begin{align}
\mathbf a_x(u)
&=
\left[
1,
e^{-j\frac{2\pi}{\lambda_c}\Delta_xu},
\ldots,
e^{-j(N_x-1)\frac{2\pi}{\lambda_c}\Delta_xu}
\right]^T,
\\
\mathbf a_y(v)
&=
\left[
1,
e^{-j\frac{2\pi}{\lambda_c}\Delta_yv},
\ldots,
e^{-j(N_y-1)\frac{2\pi}{\lambda_c}\Delta_yv}
\right]^T,
\end{align}
where \(\Delta_x\) and \(\Delta_y\) denote the center-to-center spacings between adjacent antenna elements. Let \(d_{\rm ant}\) denote the physical size of an antenna element at the transmitter, receiver, or RIS. The inter-element spacings are selected as
\(
\Delta_x=\Delta_y=\frac{5}{4}d_{\rm ant}.
\)
Accordingly, the cascaded HAPS-RIS-assisted MIMO channel is
\begin{equation}
\mathbf{F}
=
\mathbf{G}\boldsymbol{\Phi}\mathbf{H}
=
\sqrt{L_HL_G}\,
\widehat{\mathbf{G}}
\boldsymbol{\Phi}
\widehat{\mathbf{H}}
=
\sqrt{L_HL_G}\widehat{\mathbf{F}},
\label{eq:cascaded_channel}
\end{equation}
where
\(
\widehat{\mathbf{F}}
\triangleq
\widehat{\mathbf{G}}
\boldsymbol{\Phi}
\widehat{\mathbf{H}}
\)
denotes the small-scale cascaded channel.

\subsection{RIS Parameters}
The RIS is characterized by a diagonal reflection matrix
\(
\boldsymbol{\Phi}
=
\mathrm{diag}\!\big(\Gamma_{1},\ldots,\Gamma_{N_{\mathrm{RIS}}}\big),
\)
where \(\Gamma_{n}=\beta_{n}e^{j\varphi_{n}}\) with $\beta_n\in[0,1]$. Practical RIS implementations employ $b$-bit
phase shifters such that
\begin{equation}
\label{eq:phase_set}
\varphi_{n}\in\mathcal Q_\varphi=\left\{\frac{2\pi \ell}{2^b}\right\}_{\ell=0}^{2^b-1}.
\end{equation}
To account for hardware impairments, the phase-dependent amplitude
response is modeled as~\cite{RIS-AMP-RES}
\begin{equation}
\label{eq:beta_phase_dep}
\beta_{n}
=
(1-\zeta_{\min})
\left(\frac{\sin(\varphi_{n}-c)+1}{2}\right)^k+\zeta_{\min},
\end{equation}
where \(\zeta_{\min}\) denotes the minimum reflection-amplitude response, while \(c\) and \(k\) are hardware-dependent parameters.

\section{Performance Analysis}
\label{Sec:performance_analysis}

In this section, we characterize the received SNR statistics of the considered system. Since the channel model in
Section~\ref{Sec:sys_model} separates the large-scale HAPS propagation effects from the small-scale Rician fading components, the analysis is
developed in two stages. First, the distribution of the small-scale effective channel power is characterized by using a central-limit theorem (CLT)-based approximation
for the cascaded RIS channel. Then, the large-scale path-loss and log-normal shadowing terms are incorporated at the SNR level through the
multiplicative gain $L_HL_G$.

\subsection{Small-Scale Fading Statistics}
\label{subsec:small_scale_statistics}

\begin{lemma}[CLT-Based Approximation of the Cascaded Rician--Rician Channel]
\label{lem:CLT_Rician_Heff}

Consider the small-scale cascaded channel
\begin{equation}
\widehat{\mathbf F}
=
\widehat{\mathbf G}
\boldsymbol{\Phi}
\widehat{\mathbf H},
\end{equation}
where $\widehat{\mathbf H}$ and $\widehat{\mathbf G}$ independently
follow the Rician channel models in
\eqref{eq:rician_H}--\eqref{eq:rician_G}. Assuming spatially
uncorrelated NLoS components and a sufficiently large number of RIS
elements, each entry
$\widehat f_{r,t}=[\widehat{\mathbf F}]_{r,t}$ can be approximated as
\begin{equation}
\widehat f_{r,t}
=
\sum_{n=1}^{N_{\mathrm{RIS}}}
\widehat g_{r,n}\Gamma_n\widehat h_{n,t}
\approx
\mathcal{CN}
\left(
\widehat{\mu}_{r,t},
\widehat{\Sigma}
\right).
\end{equation}

The corresponding mean is
\begin{equation}
\label{eq:hat_mu_closed_final}
\widehat{\mu}_{r,t}
=
\sqrt{
\frac{K_HK_G}
{(K_H+1)(K_G+1)}
}
\eta_{\mathrm{RIS}}
\big[\mathbf a_{\mathrm R}(\vartheta_r)\big]_r
\big[\mathbf a_{\mathrm T}(\phi_t,\theta_t)\big]_t^{*},
\end{equation}
where
\(
\eta_{\mathrm{RIS}}
=
\mathbf a_{\mathrm{RIS}}^{H}(\phi_G,\theta_G)
\boldsymbol{\Phi}
\mathbf a_{\mathrm{RIS}}(\phi_H,\theta_H)
\)
denotes the coherent RIS combining factor.
For unit-variance NLoS components, i.e.,
$\sigma_h^2=\sigma_g^2=1$, the variance is identical for all
$(r,t)$ and is given by
\begin{equation}
\label{eq:hat_Sigma_closed_final}
\widehat{\Sigma}
=
\frac{K_H+K_G+1}
{(K_H+1)(K_G+1)}
\sum_{n=1}^{N_{\mathrm{RIS}}}\beta_n^2.
\end{equation}
\end{lemma}

\begin{proof}
See Appendix~\ref{app:CLT_derivation}.
\end{proof}

Lemma~\ref{lem:CLT_Rician_Heff} reveals two distinct effects of the RIS
configuration on the cascaded Rician--Rician channel. The deterministic
LoS contribution is governed by the coherent combining factor
$\eta_{\mathrm{RIS}}$, which depends on both the RIS phases and their
phase-dependent amplitudes. In contrast, the scattered contribution is
controlled by the aggregate reflected power
$\sum_{n=1}^{N_{\mathrm{RIS}}}\beta_n^2$. Hence, the RIS configuration
affects not only the deterministic channel gain but also the variance of
the equivalent small-scale channel.

\subsection{LoS-Aligned Single-Stream Beamforming and Received SNR}
\label{subsec:los_beamforming_snr}

In general, the optimal transmit beamformer maximizes the received signal
power as~\cite{5G_Precoding_Survey}
\begin{equation}
\label{eq:wopt_general}
\mathbf w_{\mathrm opt}
=
\arg\max_{\|\mathbf w\|=1}
\|\widehat{\mathbf F}\mathbf w\|^2.
\end{equation}
However, the exact solution of \eqref{eq:wopt_general} requires
instantaneous CSI and leads to an intractable eigenvalue-based statistical
characterization. In the considered HAPS-assisted aerial-to-ground
scenario, the cascaded channel is expected to exhibit a strong LoS
structure due to the elevated platform and favorable propagation geometry.
Therefore, to retain both physical consistency and mathematical
tractability, we adopt a LoS-aligned statistical precoder and set
\begin{equation}
\label{eq:precoder_mt}
\mathbf w
=
\frac{\mathbf m_t}{\|\mathbf m_t\|},
\end{equation}
where 
\(
\mathbf m_t \triangleq \mathbf a_{\mathrm T}(\phi_t,\theta_t).
\)

From Lemma~\ref{lem:CLT_Rician_Heff}, the small-scale cascaded channel can
be decomposed as
\(
\widehat{\mathbf F}
=
\widehat{\mathbf M}
+
\widehat{\mathbf X},
\)
where the deterministic mean matrix is
\begin{equation}
\widehat{\mathbf M}
=
\alpha \mathbf m_r\mathbf m_t^H,
\end{equation}
where \(\mathbf m_r\triangleq \mathbf a_{\rm R}(\phi_r,\theta_r)\), with
\begin{equation}
\label{eq:alpha_def_again}
\alpha
=
\sqrt{\frac{K_HK_G}{(K_H+1)(K_G+1)}}\eta_{\mathrm{RIS}},
\end{equation}
and $\widehat{\mathbf X}$ is the zero-mean random component. Under the LoS-aligned beamformer in \eqref{eq:precoder_mt}, the
small-scale effective receive-side channel becomes
\begin{align}
\label{eq:hat_heff_final}
\widehat{\mathbf h}_{\mathrm{eff}}
&=
\widehat{\mathbf F}\mathbf w
\nonumber\\
&=
(\widehat{\mathbf M}+\widehat{\mathbf X})\mathbf w
\nonumber\\
&=
\alpha\|\mathbf m_t\|\mathbf m_r
+
\widehat{\mathbf X}\mathbf w.
\end{align}
Defining
\begin{equation}
\label{eq:z_and_mu}
\mathbf z
\triangleq
\widehat{\mathbf X}\mathbf w,
\qquad
\boldsymbol{\mu}
\triangleq
\alpha\|\mathbf m_t\|\mathbf m_r,
\end{equation}
the effective channel can be written compactly as
\(
\widehat{\mathbf h}_{\mathrm{eff}}
=
\boldsymbol{\mu}
+
\mathbf z.
\)
Accordingly, the effective small-scale channel power is defined as
\begin{equation}
\label{eq:Q_def}
Q
\triangleq
\|\widehat{\mathbf h}_{\mathrm{eff}}\|^2
=
\|\boldsymbol{\mu}+\mathbf z\|^2.
\end{equation}

Let $s$ denote a unit-power transmitted symbol, i.e.,
$\mathbb E[|s|^2]=1$, and let $P_T$ denote the transmit power.
Using
$\mathbf F=\sqrt{L_HL_G}\widehat{\mathbf F}$,
the received signal is expressed as
\begin{equation}
\label{eq:received_signal_large_scale}
\mathbf y
=
\sqrt{P_TL_HL_G}\,
\widehat{\mathbf h}_{\mathrm{eff}}s
+
\mathbf n,
\end{equation}
where
$\mathbf n\sim\mathcal{CN}(\mathbf 0,\sigma^2\mathbf I_{N_r})$.
Assuming MRC at the receiver, the resulting post-combining SNR is
\begin{equation}
\label{eq:rho_large_scale}
\rho
=
\frac{P_T}{\sigma^2}
L_HL_G
\|\widehat{\mathbf h}_{\mathrm{eff}}\|^2
=
\frac{P_T}{\sigma^2}
L_HL_GQ.
\end{equation}
The adopted beamformer therefore exploits the deterministic transmit-side
LoS direction while avoiding instantaneous eigenmode computation. The
resulting single-stream formulation reduces the cascaded MIMO channel to
the effective random vector $\widehat{\mathbf h}_{\mathrm{eff}}$, whose
power $Q$ constitutes the key random quantity in the subsequent
performance analysis.

\subsection{Statistical Characterization of the Effective Channel Power}
\label{subsec:spa_Q}

From \eqref{eq:z_and_mu}, the effective small-scale channel is given by
\(
\widehat{\mathbf h}_{\mathrm{eff}}
=
\boldsymbol{\mu}+\mathbf z
\),
where $\mathbf z$ is approximately complex Gaussian. Hence,
\begin{equation}
\mathbf z
\sim
\mathcal{CN}
\left(
\mathbf 0,
\mathbf R_{\mathrm{eff}}
\right),
\end{equation}
where
\begin{equation}
\mathbf R_{\mathrm{eff}}
=
\mathbb E
\left[
\mathbf z\mathbf z^H
\right]
=
\mathbb E
\left[
\widehat{\mathbf X}
\mathbf w\mathbf w^H
\widehat{\mathbf X}^H
\right].
\end{equation}
Under the considered Rician--Rician channel model and the LoS-aligned
beamformer, the effective covariance matrix admits the structured form
\begin{equation}
\label{eq:Reff_def}
\mathbf R_{\mathrm{eff}}
=
a\mathbf I_{N_r}
+
b\mathbf m_r\mathbf m_r^H,
\end{equation}
where
\begin{subequations}
\begin{align}
a
&=
\xi_{\beta}
\frac{
\sigma_g^2
\left(
K_H\|\mathbf m_t\|^2+\sigma_h^2
\right)
}{
(K_G+1)(K_H+1)
},
\\
b
&=
\xi_{\beta}
\frac{
\sigma_h^2K_G
}{
(K_H+1)(K_G+1)
},
\end{align}
\end{subequations}
with
\begin{equation}
\label{eq:xi_beta}
\xi_{\beta}
=
\sum_{n=1}^{N_{\mathrm{RIS}}}\beta_n^2.
\end{equation}
\begin{proof}
See 
Appendix~\ref{app:Reff_derivation}.
\end{proof}
Since \eqref{eq:Reff_def} is a rank-one perturbation of a scaled identity
matrix, its eigenstructure can be obtained explicitly. Let
\begin{equation}
\label{eq:Reff_eigendecomposition}
\mathbf R_{\mathrm{eff}}
=
\mathbf U
\boldsymbol{\Lambda}
\mathbf U^H,
\end{equation}
where $\mathbf U$ is a unitary matrix whose first column is aligned with
$\mathbf m_r$, i.e.,
\begin{equation}
\mathbf u_{\parallel}
=
\frac{\mathbf m_r}{\|\mathbf m_r\|}.
\end{equation}
Accordingly, the eigenvalue associated with the direction
$\mathbf u_{\parallel}$ is
\begin{equation}
\label{eq:lambda_parallel}
\lambda_{\parallel}
=
a+b\|\mathbf m_r\|^2,
\end{equation}
whereas the remaining $N_r-1$ orthogonal eigenmodes have the repeated
eigenvalue
\(
\lambda_{\perp}
=
a.
\)
Hence, the eigenvalue matrix can be written as
\begin{equation}
\boldsymbol{\Lambda}
=
\mathrm{diag}
\left(
\lambda_{\parallel},
\lambda_{\perp},
\ldots,
\lambda_{\perp}
\right).
\end{equation}
Applying the unitary transformations
\(
\widetilde{\mathbf z}
=
\mathbf U^H\mathbf z
\)
and
\(
\widetilde{\boldsymbol{\mu}}
=
\mathbf U^H\boldsymbol{\mu},
\)
we obtain
\(
\widetilde{\mathbf z}
\sim
\mathcal{CN}(\mathbf 0,\boldsymbol{\Lambda})
\).
Since $\boldsymbol{\Lambda}$ is diagonal, the components of
$\widetilde{\mathbf z}$ are mutually independent. Moreover, the
deterministic mean vector is given by
\begin{equation}
\boldsymbol{\mu}
=
\alpha
\|\mathbf m_t\|
\mathbf m_r,
\end{equation}
and is therefore aligned with the direction
$\mathbf u_{\parallel}$. Consequently,
\begin{equation}
\widetilde{\boldsymbol{\mu}}
=
\begin{bmatrix}
\mu_{\parallel} &
0 &
\cdots &
0
\end{bmatrix}^{T},
\end{equation}
where
\(
\label{eq:mu_parallel}
\mu_{\parallel}
=
\alpha
\|\mathbf m_t\|
\|\mathbf m_r\|.
\)
Using the unitary invariance of the Euclidean norm, the effective
small-scale channel power can now be written as
\begin{align}
Q
&=
\left\|
\boldsymbol{\mu}
+
\mathbf z
\right\|^2
=
\left\|
\widetilde{\boldsymbol{\mu}}
+
\widetilde{\mathbf z}
\right\|^2
\nonumber\\
&=
\left|
\mu_{\parallel}
+
\widetilde z_{\parallel}
\right|^2
+
\sum_{i=2}^{N_r}
\left|
\widetilde z_i
\right|^2,
\label{eq:Q_parallel_perp}
\end{align}
where
\(
\widetilde z_{\parallel}
\sim
\mathcal{CN}
\left(
0,\lambda_{\parallel}
\right),
\)
and
\(
\widetilde z_i
\sim
\mathcal{CN}
\left(
0,\lambda_{\perp}
\right),
\)
with \(i=2,\ldots,N_r\).
The random variables $\widetilde z_{\parallel},\widetilde z_2,\ldots,
\widetilde z_{N_r}$ are mutually independent. Hence, \eqref{eq:Q_parallel_perp} shows that $Q$ consists of one noncentral quadratic component associated with the dominant LoS eigenmode and $N_r-1$ central components associated with the orthogonal scattering
modes. Since these components generally have different scale parameters, $Q$ does not reduce to a single Gamma random variable.
Therefore, we employ SPA to obtain compact approximations of its PDF and CDF from the corresponding CGF. 
\begin{lemma}[SPA-Based PDF and CDF of the Effective Channel Power]
\label{lem:SPA_Q}

From \eqref{eq:Q_parallel_perp}, define \(Y_{\parallel}\triangleq\left|\mu_{\parallel}+\widetilde z_{\parallel}\right|^2\) and \(Y_i\triangleq|\widetilde z_i|^2\), \(i=2,\ldots,N_r\). Accordingly,
\begin{equation}
\label{eq:Q_sum_components}
Q
=
Y_{\parallel}
+
\sum_{i=2}^{N_r}Y_i.
\end{equation}
Since
\(
\widetilde z_{\parallel}
\sim
\mathcal{CN}(0,\lambda_{\parallel})
\),
the random variable $Y_{\parallel}$ follows a scaled noncentral
chi-square distribution with two degrees of freedom. Its MGF is given by
\begin{equation}
\label{eq:MGF_Y_parallel}
M_{Y_{\parallel}}(s)
=
\frac{1}{1-s\lambda_{\parallel}}
\exp\left(
\frac{s|\mu_{\parallel}|^2}
{1-s\lambda_{\parallel}}
\right),
\qquad
s<\lambda_{\parallel}^{-1}.
\end{equation}
On the other hand, since
\(
\widetilde z_i
\sim
\mathcal{CN}(0,\lambda_{\perp})
\)
for $i=2,\ldots,N_r$, each
\(
Y_i=|\widetilde z_i|^2
\)
is exponentially distributed with mean $\lambda_{\perp}$.
Hence, its MGF is
\begin{equation}
\label{eq:MGF_Y_perp}
M_{Y_i}(s)
=
\frac{1}{1-s\lambda_{\perp}},
\qquad
s<\lambda_{\perp}^{-1}.
\end{equation}
Since
$Y_{\parallel},Y_2,\ldots,Y_{N_r}$
are mutually independent, the MGF of their sum in
\eqref{eq:Q_sum_components} is obtained as the product of the individual
MGFs:
\begin{align}
M_Q(s)
&=
M_{Y_{\parallel}}(s)
\prod_{i=2}^{N_r}
M_{Y_i}(s)
\nonumber\\
&=
\frac{
\exp\left(
\frac{s|\mu_{\parallel}|^2}
{1-s\lambda_{\parallel}}
\right)
}{
(1-s\lambda_{\parallel})
(1-s\lambda_{\perp})^{N_r-1}
},
\label{eq:MGF_Q_structured}
\end{align}
for
\(
s<
\min
\left\{
\lambda_{\parallel}^{-1},
\lambda_{\perp}^{-1}
\right\}.
\)
Accordingly, the CGF of $Q$ becomes
\begin{align}
\label{eq:CGF_Q_structured}
K_Q(s)
&\triangleq
\ln M_Q(s)
\nonumber\\
&=
-\ln(1-s\lambda_{\parallel})
-(N_r-1)\ln(1-s\lambda_{\perp})
\nonumber\\
&\quad+
\frac{s|\mu_{\parallel}|^2}
{1-s\lambda_{\parallel}}.
\end{align}
Its first and second derivatives are respectively given by
\begin{subequations}
\begin{align}
&K_Q'(s)=
\frac{\lambda_{\parallel}}
{1-s\lambda_{\parallel}}
+
\frac{(N_r-1)\lambda_{\perp}}
{1-s\lambda_{\perp}}+
\frac{|\mu_{\parallel}|^2}
{(1-s\lambda_{\parallel})^2},
\label{eq:CGF_Q_prime_structured}
\\
&K_Q''(s)
=
\frac{\lambda_{\parallel}^2}
{(1-s\lambda_{\parallel})^2}
+
\frac{(N_r-1)\lambda_{\perp}^2}
{(1-s\lambda_{\perp})^2} +
\frac{
2\lambda_{\parallel}|\mu_{\parallel}|^2
}{
(1-s\lambda_{\parallel})^3
}.
\label{eq:CGF_Q_second_structured}
\end{align}
\end{subequations}
For a given $q>0$, the saddlepoint $\hat{s}$ is obtained as the solution
of
\(
K_Q'(\hat{s})=q.
\)
The SPA-based PDF of $Q$ is then approximated as~\cite{daniels1954saddlepoint}
\begin{equation}
\label{eq:pdf_Q_spa}
f_Q(q)
\approx
\frac{
\exp\left(
K_Q(\hat{s})-\hat{s}q
\right)
}{
\sqrt{
2\pi K_Q''(\hat{s})
}
}.
\end{equation}
For the corresponding CDF, define
\begin{subequations}
\begin{align}
w
&=
\mathrm{sign}(\hat{s})
\sqrt{
2\left(
\hat{s}q-K_Q(\hat{s})
\right)
},
\\
u
&=
\hat{s}
\sqrt{
K_Q''(\hat{s})
}.
\end{align}
\end{subequations}
Using the Lugannani--Rice approximation, the CDF is obtained as
\begin{equation}
\label{eq:cdf_Q_spa}
F_Q(q)
\approx
\Psi(w)
+
\psi(w)
\left(
\frac{1}{w}
-
\frac{1}{u}
\right),
\end{equation}
where $\Psi(\cdot)$ and $\psi(\cdot)$ denote the standard Gaussian CDF
and PDF, respectively.
\end{lemma}

 \subsection{Received SNR Distribution and Outage Analysis}
\label{subsec:SNR_distribution}

Having characterized the PDF and CDF of the effective small-scale
channel power $Q$, we next derive the statistical distribution of the
received SNR by incorporating the large-scale propagation effects.
Define the composite large-scale channel gain as
\(
L_{\mathrm{tot}}
\triangleq
L_HL_G.
\)
From \eqref{eq:rho_large_scale}, the received SNR can then be expressed as
\begin{equation}
\label{eq:rho_compact}
\rho
=
\frac{P_T}{\sigma^2}
L_{\mathrm{tot}}Q.
\end{equation}
For a given realization $L_{\mathrm{tot}}=l$, the received SNR is a
scaled version of $Q$. Hence,
\begin{subequations}
\label{eq:snr_conditional}
\begin{align}
f_{\rho|L_{\mathrm{tot}}}(\rho|l)
&=
\frac{\sigma^2}{P_Tl}
f_Q\left(
\frac{\rho\sigma^2}{P_Tl}
\right),
\label{eq:snr_conditional_pdf}
\\
F_{\rho|L_{\mathrm{tot}}}(\rho|l)
&=
F_Q\left(
\frac{\rho\sigma^2}{P_Tl}
\right).
\label{eq:snr_conditional_cdf}
\end{align}
\end{subequations}
Assuming independence between the small-scale fading and large-scale
shadow fading, the unconditional SNR distribution is obtained by
averaging over the random composite gain $L_{\mathrm{tot}}$.
From \eqref{eq:L_total_large_scale}, the composite large-scale gain is
given by
\begin{equation}
\label{eq:Ltot_def_final}
L_{\mathrm{tot}}
=
\mathcal{L}_H\mathcal{L}_G
10^{-\frac{
\chi_H+\chi_G+
PL_{A,H}(p)+PL_{A,G}(p)
}{10}}.
\end{equation}
Since $\chi_H$ and
$\chi_G$ are assumed to be independent Gaussian random variables in
the dB domain, their sum is also Gaussian. Consequently,
$L_{\mathrm{tot}}$ remains log-normally distributed in the linear
domain and can equivalently be represented as
\begin{equation}
\label{eq:Ltot_lognormal}
L_{\mathrm{tot}}
=
\exp\left(
\mu_L(p)+\sigma_L X
\right),
\end{equation}
where
\(
X\sim\mathcal{N}(0,1),
\)
with
\begin{equation}
\label{eq:mu_L}
\mu_L(p)
=
\ln\left(
\mathcal{L}_H\mathcal{L}_G
\right)
-
\frac{\ln 10}{10}
\left[
PL_{A,H}(p)+PL_{A,G}(p)
\right],
\end{equation}
and
\begin{equation}
\label{eq:sigma_L}
\sigma_L
=
\frac{\ln 10}{10}
\sqrt{
\sigma_{\chi,H}^{2}
+
\sigma_{\chi,G}^{2}
}.
\end{equation}
Accordingly, the unconditional PDF and CDF of the received SNR are
given by
\begin{subequations}
\begin{align}
\label{eq:snr_pdf_integral}
f_{\rho}(\rho)
&=\int_{0}^{\infty}
\frac{\sigma^2}{P_Tl}
f_Q\left(
\frac{\rho\sigma^2}{P_Tl}
\right)
f_{L_{\mathrm{tot}}}(l)
\,dl,
\\
\label{eq:snr_cdf_integral}
F_{\rho}(\rho)
&=
\int_{0}^{\infty}
F_Q\left(
\frac{\rho\sigma^2}{P_Tl}
\right)
f_{L_{\mathrm{tot}}}(l)
\,dl.
\end{align}
\end{subequations}
Equations~\eqref{eq:snr_pdf_integral} and
\eqref{eq:snr_cdf_integral} jointly incorporate the small-scale
cascaded Rician--Rician fading, log-normal shadow fading, and
atmospheric attenuation corresponding to the prescribed exceedance
probability $p$. Although these expressions provide one-dimensional
integral representations of the complete SNR distribution, their
repeated numerical evaluation may become computationally demanding
over a wide SNR range or for multiple system configurations.

To obtain computationally efficient expressions for the received SNR
distribution, we apply Gauss--Hermite quadrature to
\eqref{eq:snr_pdf_integral} and \eqref{eq:snr_cdf_integral} by
exploiting the Gaussian representation in \eqref{eq:Ltot_lognormal}.
Let $\{x_i,w_i\}_{i=1}^{N_{\mathrm{GH}}}$ denote the abscissas and
weights of an $N_{\mathrm{GH}}$-point Gauss--Hermite quadrature rule.
The resulting SNR PDF and CDF are approximated as \eqref{eq:snr_pdf_GH} and \eqref{eq:snr_cdf_GH}, respectively.
\begin{subequations}
\label{eq:snr_GH}
\begin{align}
f_{\rho}(\rho)
&\approx
\frac{1}{\sqrt{\pi}}
\sum_{i=1}^{N_{\mathrm{GH}}}
w_i
\frac{
\sigma^2
}{
P_T
\exp\left(
\mu_L(p)+\sqrt{2}\sigma_Lx_i
\right)
}
\nonumber\\
&\quad\times
f_Q\left(
\frac{
\rho\sigma^2
}{
P_T
\exp\left(
\mu_L(p)+\sqrt{2}\sigma_Lx_i
\right)
}
\right),
\label{eq:snr_pdf_GH}
\\[0.5em]
F_{\rho}(\rho)
&\approx
\frac{1}{\sqrt{\pi}}
\sum_{i=1}^{N_{\mathrm{GH}}}
w_i
F_Q\left(
\frac{
\rho\sigma^2
}{
P_T
\exp\left(
\mu_L(p)+\sqrt{2}\sigma_Lx_i
\right)
}
\right).
\label{eq:snr_cdf_GH}
\end{align}
\end{subequations}
\begin{proof}
See Appendix~\ref{app:GH_outage}.
\end{proof} 
These expressions provide the PDF and CDF of the received SNR by
combining the SPA-based characterization of the small-scale effective
channel power with log-normal shadow fading and the atmospheric
attenuation associated with the prescribed exceedance probability
$p$.

For a prescribed SNR threshold \(\rho_{\mathrm{th}}\), an outage event is defined as \(\rho<\rho_{\mathrm{th}}\). Hence, the corresponding outage probability follows directly from \eqref{eq:snr_cdf_GH} as
\(
P_{\mathrm{out}}(\rho_{\mathrm{th}})
=
F_{\rho}(\rho_{\mathrm{th}}).
\)
Accordingly, \eqref{eq:snr_cdf_GH} constitutes the final outage characterization of the considered HAPS-RIS-assisted multiple-antenna links for a prescribed atmospheric exceedance probability \(p\). The resulting expression jointly captures the antenna configuration, RIS size, finite phase resolution, phase-dependent amplitude response, cascaded Rician--Rician small-scale fading, log-normal shadowing, and location-dependent atmospheric attenuation. 


\subsection{LoS-Aligned RIS Phase Configuration}
\label{subsec:ris_phase_design}

Due to the dominant LoS structure and the adopted LoS-aligned
beamforming strategy, the RIS phases are configured to coherently
combine the deterministic LoS components. For the continuous-phase
case, the phase-aligned configuration of the $n$th RIS element is
selected as
\begin{equation}
\label{eq:los_aligned_continuous_phase}
\varphi_{\mathrm{c},n}
=
-\angle\!\left(
\big[\mathbf a_{\mathrm{RIS}}(\phi_G,\theta_G)\big]_n^{*}
\big[\mathbf a_{\mathrm{RIS}}(\phi_H,\theta_H)\big]_n
\right).
\end{equation}
For a $b$-bit RIS, the corresponding discrete phase is obtained by
nearest-neighbor quantization as
\begin{equation}
\label{eq:los_aligned_discrete_phase}
\varphi_{\mathrm{d},n}
=
\arg\min_{\varphi\in\mathcal Q_\varphi}
\left|
\varphi-\varphi_{\mathrm{c},n}
\right|.
\end{equation}
It should be emphasized that the configuration in
\eqref{eq:los_aligned_continuous_phase}--\eqref{eq:los_aligned_discrete_phase}
is designed to align the deterministic LoS phases and does not, in
general, constitute the globally optimal RIS configuration under the
phase-dependent amplitude response in \eqref{eq:beta_phase_dep}. Since
the reflection amplitude $\beta_n$ varies with the applied phase shift,
maximizing the overall effective channel gain would require a joint
consideration of the phase and amplitude responses.

\section{Results and Practical Insights}
\label{Sec:simulation_results}

In this section, the proposed analytical expressions are validated through Monte Carlo simulations. To incorporate realistic environment-dependent large-scale
propagation effects, the atmospheric attenuation components are
evaluated using the corresponding ITU-R models for the selected
geographical locations. As shown in Fig.~\ref{fig:Sim_mod_Google}, the
transmitter is located at TÜBİTAK in Kocaeli Province, while the
receiver is located at Sabiha Gökçen International Airport (SAW) in Istanbul
Province. The corresponding HAPS-RIS-assisted deployment geometry is
illustrated in Fig.~\ref{fig:Sim_mod_Geometry}.

\subsection{Deployment Geometry}
For the geometry in Fig.~\ref{fig:Sim_mod_Geometry}, the HAPS-RIS
center is selected as the origin of the global Cartesian coordinate
system,
\(
\mathbf p_{\mathrm{RIS}}=[0,0,0]^T.
\)
The global axes $x_g$, $y_g$, and $z_g$ are aligned with the east,
north, and zenith directions, respectively, following the orientation of the local ENU frame defined at the HAPS nadir point.

\begin{figure}[t!]
    \centering {\includegraphics[width=3.4in, angle=0]{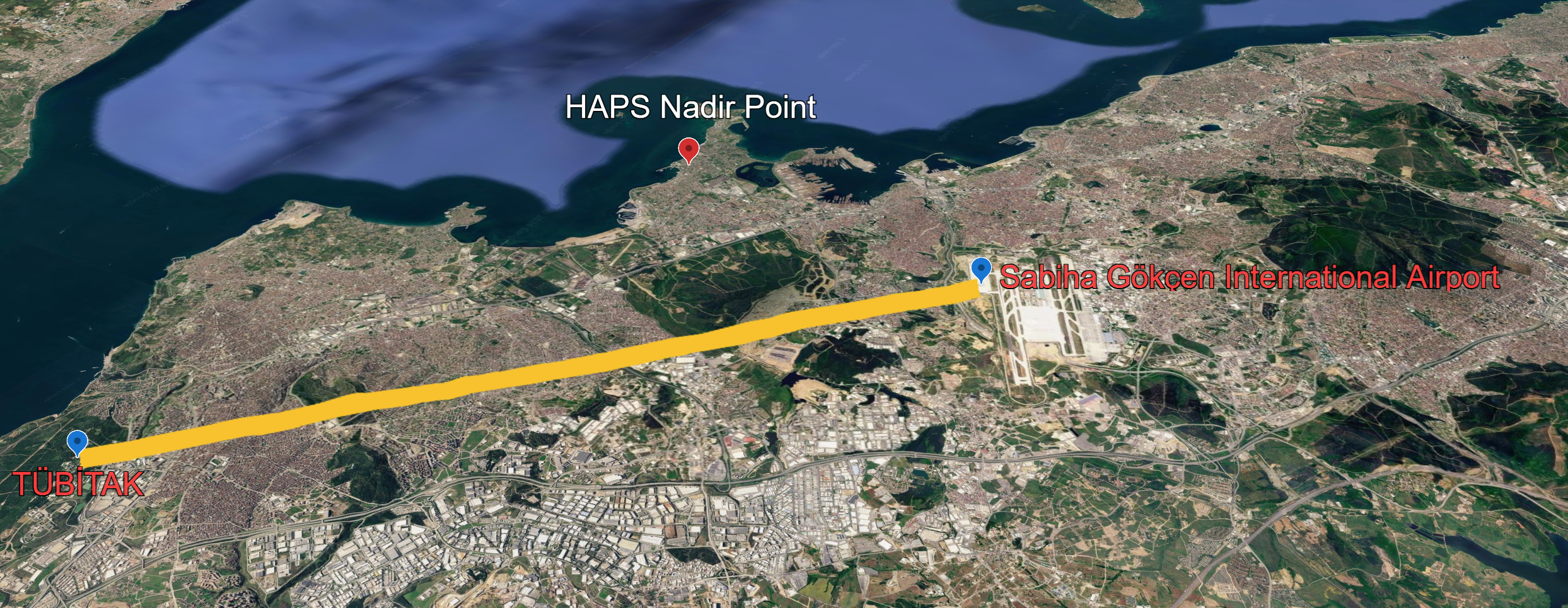}}
    \caption{Considered ground-station locations for the simulation setup, where the transmitter is placed at TÜBİTAK and the receiver at Sabiha Gökçen International Airport.}
\label{fig:Sim_mod_Google}
\end{figure}

\begin{figure}[t!]
    \centering {\includegraphics[width=\columnwidth, angle=0]{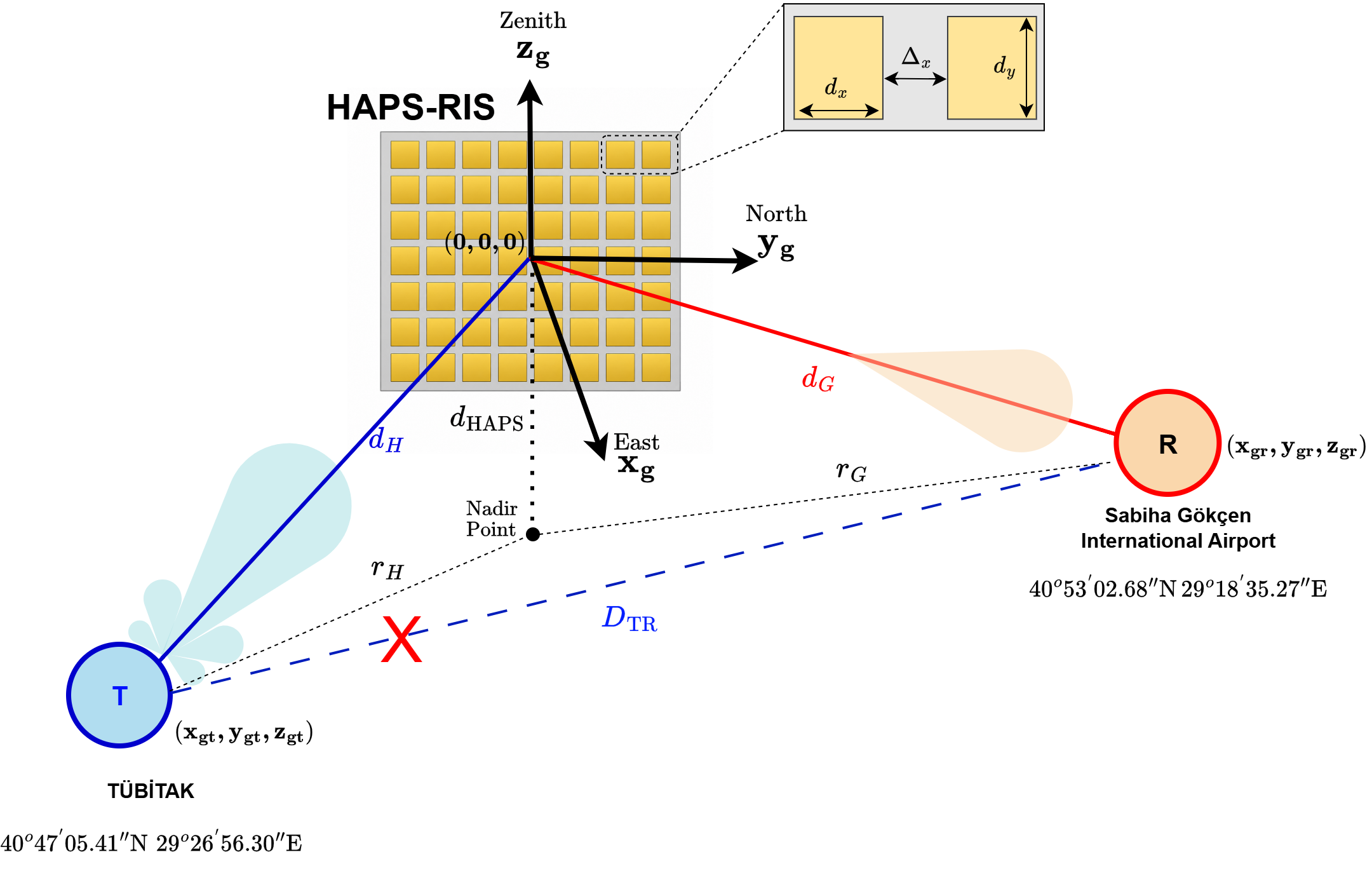}}
    \caption{Geometric representation of the HAPS-RIS-assisted link corresponding to the ground-station locations in Fig.~\ref{fig:Sim_mod_Google}, showing the slant distances and angular parameters.}
\label{fig:Sim_mod_Geometry}
\end{figure}

To map the real deployment in Fig.~\ref{fig:Sim_mod_Google} to this
Cartesian geometry, the geodetic coordinates of the transmitter,
receiver, and selected HAPS nadir point are first transformed to ECEF coordinates using the WGS-84
reference ellipsoid. The resulting terminal coordinates are then
transformed to a local ENU frame referenced at the HAPS nadir
point\footnote{The WGS-84 geodetic-to-ECEF and
ECEF-to-ENU transformations are numerically implemented using the
\texttt{geodetic2ecef} and \texttt{ecef2enu} functions of the MATLAB
Mapping Toolbox.}.
Neglecting the relatively small difference between the ground
elevations of the two locations, the transmitter and receiver
positions are written directly in terms of their ENU offsets as
\(
\mathbf p_T
=
[E_T,N_T,-d_{\mathrm{HAPS}}]^T,
\) \(\mathbf p_R
=
[E_R,N_R,-d_{\mathrm{HAPS}}]^T,
\)
where $d_{\mathrm{HAPS}}$ denotes the vertical distance between the
HAPS-RIS center and its nadir point on the ground.
The horizontal ranges from the HAPS nadir point are
\begin{equation}
r_H
=
\sqrt{E_T^2+N_T^2},
\qquad
r_G
=
\sqrt{E_R^2+N_R^2},
\label{eq:horizontal_ranges}
\end{equation}
and the corresponding T--RIS and RIS--R slant distances are
\begin{equation}
d_H
=
\sqrt{r_H^2+d_{\mathrm{HAPS}}^2},
\qquad
d_G
=
\sqrt{r_G^2+d_{\mathrm{HAPS}}^2}.
\label{eq:slant_distances_sim}
\end{equation}
For the ITU-R atmospheric attenuation calculations, the physical link elevation angles, measured from the local horizontal plane, are
\(
\alpha_H
=
\operatorname{atan2}(d_{\mathrm{HAPS}},r_H)
\)
and
\(
\alpha_G
=
\operatorname{atan2}(d_{\mathrm{HAPS}},r_G),
\)
where \(\operatorname{atan2}(\cdot,\cdot)\) denotes the four-quadrant inverse tangent function. The geodetic coordinates, resulting ENU
offsets, and corresponding geometric link parameters for the
considered deployment are summarized in
Table~\ref{tab:deployment_geometry}.

The LoS angular parameters used in the array steering vectors are
obtained from the same deployment geometry. Each URA is defined in its
local $x$-$y$ plane, with the local $+z$-axis corresponding to the
array broadside direction. For the transmitter and receiver arrays,
the local $x$-, $y$-, and $z$-axes are aligned with the east, north,
and zenith directions, respectively. For the HAPS-RIS,
whose broadside is directed toward the ground, the local axes are
aligned with east, south, and nadir, respectively.
Accordingly, the LoS direction vectors expressed in the corresponding
local array frames are
\begin{align}
\mathbf v_t^{(a)}
&=
[-E_T,-N_T,d_{\mathrm{HAPS}}]^T,
&
\mathbf v_H^{(a)}
&=
[E_T,-N_T,d_{\mathrm{HAPS}}]^T,
\nonumber\\
\mathbf v_G^{(a)}
&=
[E_R,-N_R,d_{\mathrm{HAPS}}]^T,
&
\mathbf v_r^{(a)}
&=
[-E_R,-N_R,d_{\mathrm{HAPS}}]^T,
\label{eq:local_link_direction_vectors}
\end{align}
where the subscripts $t$, $H$, $G$, and $r$ correspond to the
transmitter AoD, RIS AoA, RIS AoD, and receiver AoA directions,
respectively.

For a generic local direction vector
\(\mathbf v^{(a)}=[v_x,v_y,v_z]^T\), the corresponding azimuth and polar angles are defined as
\(
\phi(\mathbf v^{(a)})
=
\operatorname{atan2}(v_y,v_x)
\)
and
\(
\theta(\mathbf v^{(a)})
=
\cos^{-1}
\left(
\frac{v_z}{\|\mathbf v^{(a)}\|}
\right).
\)
Hence,
\begin{align}
\phi_t
&=
\operatorname{atan2}(-N_T,-E_T),
&
\phi_H
&=
\operatorname{atan2}(-N_T,E_T),
\nonumber\\
\phi_G
&=
\operatorname{atan2}(-N_R,E_R),
&
\phi_r
&=
\operatorname{atan2}(-N_R,-E_R).
\label{eq:azimuth_angles_sim}
\end{align}
Under the adopted array orientations, the broadside component of each
LoS direction vector is $d_{\mathrm{HAPS}}$. Therefore, the polar
angles satisfy
\begin{align}
\theta_t=\theta_H
&=
\cos^{-1}\left(\frac{d_{\mathrm{HAPS}}}{d_H}\right)
=
\frac{\pi}{2}-\alpha_H,
\\
\theta_G=\theta_r
&=
\cos^{-1}\left(\frac{d_{\mathrm{HAPS}}}{d_G}\right)
=
\frac{\pi}{2}-\alpha_G.
\label{eq:polar_elevation_relation}
\end{align}

\subsection{Simulation Parameters and Atmospheric Attenuations}

The simulation parameters are selected consistently with the geographical deployment described in the previous subsection. The considered links are assumed to operate under suburban LOS conditions
in the Ka band. The shadow-fading standard deviations and Rician $K$-factors are selected from the corresponding 3GPP TR~38.811 suburban LOS parameter tables using the nearest reference elevation
angles to the calculated link elevations~\cite{3GPP_TR_38811}. Specifically,
$\alpha_H=55.53^\circ$ is associated with the $60^\circ$ reference angle, whereas $\alpha_G=69.16^\circ$ is associated with the
$70^\circ$ reference angle. Accordingly,
$\sigma_{\chi,H}=3.1$ dB and $\sigma_{\chi,G}=3.0$ dB are adopted for the log-normal shadowing terms, while the corresponding mean Rician factors are selected as $K_H=6.6$ dB and $K_G=5.9$ dB. The 3GPP suburban LOS Ka-band model provides these parameters explicitly as a function of the reference elevation angle~\cite{3GPP_TR_38811}.

\subsubsection{Long-Term Rain Attenuation}
The long-term rain attenuation is evaluated for the specific
geographical coordinates of the considered T--RIS and RIS--R links
by following the prediction procedure in Section~2.2.1.1 of
ITU-R P.618-14~\cite{ITU_R_P618}. In this procedure, the point
rainfall rate $R_{0.01}$, exceeded for $0.01\%$ of an average year,
is obtained from the digital maps provided with ITU-R
P.837-8~\cite{ITU_R_P837}. The corresponding digital data are
available in the supplementary files of the Recommendation.
For the considered geographical locations, bilinear interpolation of
the $R_{0.01}$ map yields
$R_{0.01,H}=30.6729$ mm/h and
$R_{0.01,G}=30.5189$ mm/h.
The rain height is determined according to ITU-R
P.839-4~\cite{ITU_R_P839} as
$h_R=h_0+0.36$ km, where $h_0$ denotes the mean annual
$0^\circ$C isotherm height above mean sea level. The location-specific
values of $h_0$ are obtained from the digital maps accompanying
ITU-R P.839-4.
This results in $h_{R,H}=2.9825$ km and
$h_{R,G}=2.9484$ km for the two links.

Using these location-dependent parameters together with the
link-specific elevation angles, the carrier frequency, and the
rain-specific attenuation coefficients prescribed by ITU-R
P.838-3~\cite{ITU_R_P838}, the rain attenuation
$A_{R,\ell}(p)$ is calculated according to the remaining steps of
ITU-R P.618-14. Horizontal linear polarization is assumed in the
numerical evaluation. For brevity, the intermediate calculations are
not reproduced here. The resulting rain attenuation values for the
considered annual exceedance probabilities are summarized in
Table~\ref{tab:atmospheric_attenuation}.
\begin{table}[t]
\centering
\caption{Geographical and geometric parameters of the considered
deployment.}
\label{tab:deployment_geometry}
\begin{tabular}{lc}
\hline
Parameter & Value \\
\hline
HAPS nadir coordinates
& $(40^\circ49'00.76''\mathrm{N},\,
29^\circ17'30.63''\mathrm{E})$ \\

TÜBİTAK coordinates
& $(40^\circ47'05.41''\mathrm{N},\,
29^\circ26'56.30''\mathrm{E})$ \\

SAW coordinates
& $(40^\circ53'02.68''\mathrm{N},\,
29^\circ18'35.27''\mathrm{E})$ \\

$d_{\mathrm{HAPS}}$
& $20$ km \\

$(E_T,N_T)$
& $(13.2631,\,-3.5463)$ km \\

$(E_R,N_R)$
& $(1.5133,\,7.4628)$ km \\

$r_H$,~$r_G$
& $13.7290$ km,~$7.6147$ km \\

$d_H$,~$d_G$
& $24.2587$ km,~$21.4005$ km \\

$D_{\mathrm{TR}}$
& $16.1015$ km \\

$\alpha_H$,~$\alpha_G$
& $55.5323^\circ$,~$69.1565^\circ$  \\
\hline
\end{tabular}
\end{table}

\subsubsection{Tropospheric Scintillation Attenuation}

The tropospheric scintillation attenuation is evaluated according to
Section~2.4.1 of ITU-R P.618-14~\cite{ITU_R_P618}. Since the
free-space elevation angles of both considered links exceed $5^\circ$,
the long-term scintillation prediction procedure for high-elevation
Earth-space links is applied. The annual median wet term of the
surface refractivity, $N_{\rm wet}$, is obtained from the digital maps
provided with ITU-R P.453~\cite{ITU_R_P453} and bilinearly interpolated
at the geographical coordinates of the two ground terminals.
Using the corresponding $N_{\rm wet}$ values together with the carrier
frequency, link elevation angles, physical terminal antenna diameter,
and antenna efficiency, the scintillation attenuation
$A_{S,\ell}(p)$ is calculated following the steps prescribed in
ITU-R P.618-14. 

For the tropospheric scintillation calculation, the physical ground-terminal antenna diameter is set to
\(D=142\lambda_c\), corresponding to approximately \(1.37\)~m at
\(f_c=31\)~GHz. The antenna efficiency is set to \(\eta=0.5\), consistent with the conservative value recommended in ITU-R P.618-14 when the efficiency is not otherwise specified. For brevity, the intermediate calculations are not reproduced here. The resulting values are summarized in Table~\ref{tab:atmospheric_attenuation}.

\begin{figure*}[t]
    \centering

    \begin{minipage}[c]{0.32\textwidth}
        \centering
        \includegraphics[width=\linewidth]{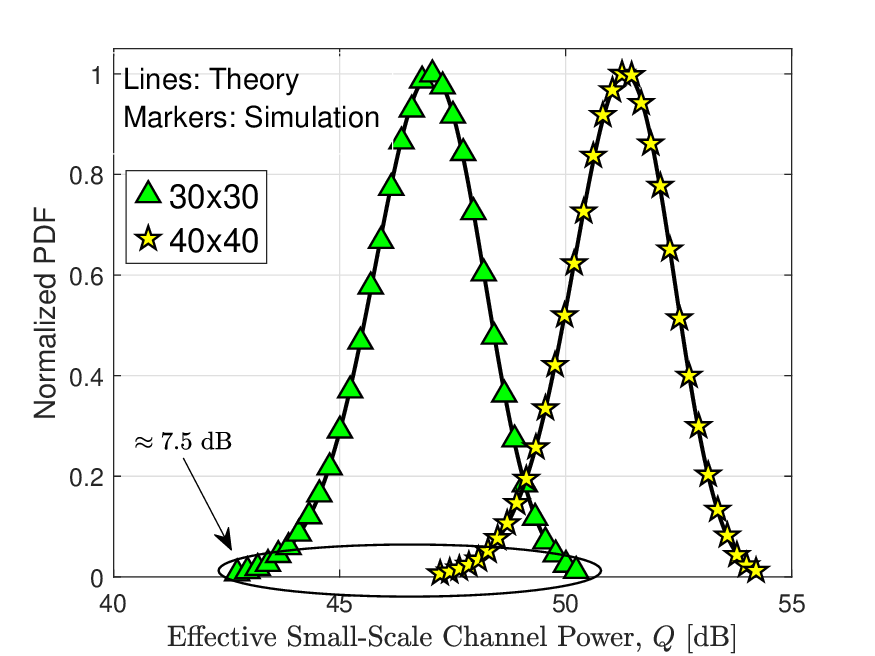}

        \small (a) Random phase assignment
    \end{minipage}
    \hfill
    \begin{minipage}[c]{0.32\textwidth}
        \centering
        \includegraphics[width=\linewidth]{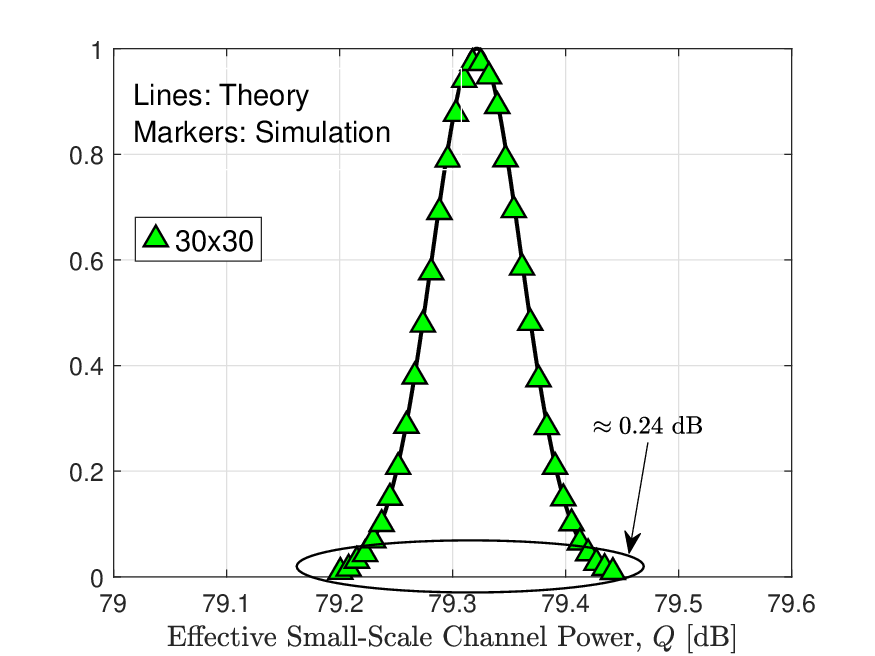}

        \small (b) LoS-aligned, $30\times30$
    \end{minipage}
    \hfill
    \begin{minipage}[c]{0.32\textwidth}
        \centering
        \includegraphics[width=\linewidth]{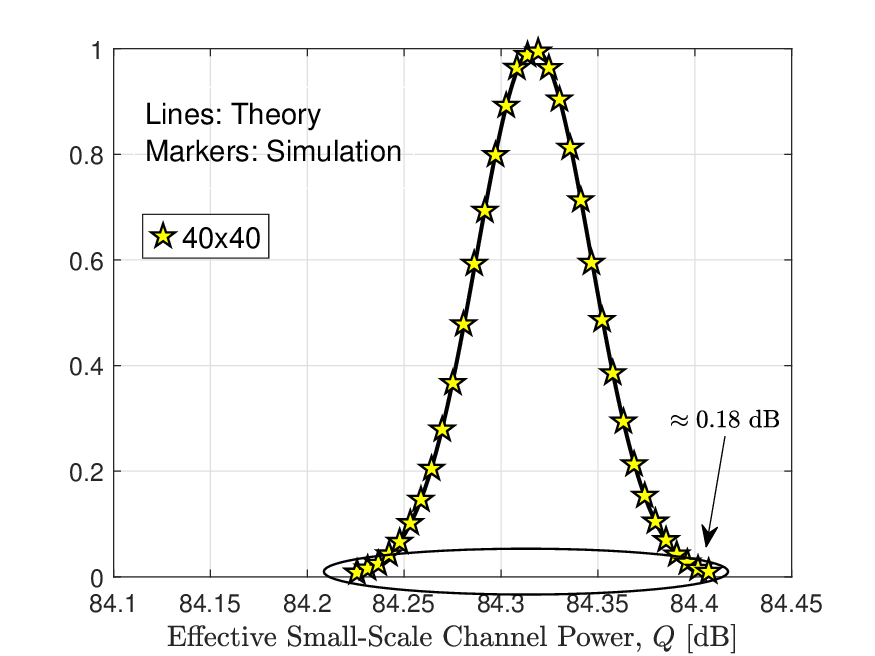}

        \small (c) LoS-aligned, $40\times40$
    \end{minipage}

    \caption{Normalized PDF of the effective small-scale channel power
$Q$ for random and LoS-aligned RIS phase configurations~\((b=2)\).}
    \label{fig:q_pdf_validation}
\end{figure*}

\subsubsection{Gaseous Attenuation}
The gaseous attenuation is evaluated according to ITU-R
P.676-13~\cite{ITU_R_P676}, which provides statistical prediction
methods for the oxygen and water-vapour contributions to the
slant-path gaseous attenuation. The meteorological parameters required
by this procedure are obtained from the annual digital maps provided
with ITU-R P.2145-0~\cite{ITU_R_P2145}. In particular, the surface
barometric pressure, surface temperature, surface water-vapour
density, and integrated water-vapour content statistics are extracted
for the geographical coordinates of the two ground terminals. The
corresponding values are obtained by applying the height-scaling and
bilinear spatial interpolation procedures specified in
ITU-R P.2145-0.
The resulting meteorological parameters are then used in the
statistical gaseous-attenuation procedure of Annex~2 of
ITU-R P.676-13 to determine the total gaseous attenuation
$A_{G,\ell}(p)$ for each link. The resulting values are summarized in
Table~\ref{tab:atmospheric_attenuation}.

\subsubsection{Cloud Attenuation}

The cloud attenuation is evaluated according to the statistical
slant-path prediction method of ITU-R P.840-9~\cite{ITU_R_P840}.
The annual integrated cloud liquid water content statistics required by this procedure are obtained from the digital maps provided with ITU-R P.840-9 for the geographical coordinates of the two ground
terminals. The corresponding values are determined using the bilinear spatial interpolation procedure specified in the Recommendation.
The cloud liquid mass absorption coefficient at $f_c=31$~GHz is
calculated according to the frequency-dependent model of
ITU-R P.840-9. The resulting
values are summarized in Table~\ref{tab:atmospheric_attenuation}.
\begin{table}[t]
\centering
\caption{Other Simulation parameters.}
\label{tab:simulation_parameters}
\begin{tabular}{lc}
\hline
Parameter & Value \\
\hline
Carrier frequency, $f_c$ & $31$ GHz \\
Propagation environment & Suburban LOS \\
Tx/Rx antennas, $N_t\times N_r$ & $16\times 16$ \\
Tx/Rx antenna-element size, $d_{\rm ant}$ & $142\lambda_c$ \\
Tx/Rx antenna-element gains, $G_t,G_r$ & $50$ dBi \\
RIS size, $N_{\mathrm{RIS}}$ & $30^2,\;40^2,\;50^2$ \\
Unit RIS element size & $\lambda_c/2$ \\
$\sigma_{\chi,H},\,\,\sigma_{\chi,G}$ & $3.1,\,\,3.0$ dB \\
$K_H,\,\,K_G$ & $6.6,\,\,5.9$ dB\\
Noise power, $\sigma^2$ & $-114$ dBm \\
RIS parameters, $\zeta_{\min}$, $c$, $k$ & $0.8,~0.43\pi,~1.6$\\
\hline
\end{tabular}
\end{table}

\begin{table}[t]
\centering
\caption{Location-specific atmospheric attenuation components and total attenuation for the T--RIS (\(\ell=H\)) and RIS--R (\(\ell=G\)) links.}
\label{tab:atmospheric_attenuation}
\begin{tabular}{c c|c c c c|c}
\hline
\(p\) (\%) & Hop &
\(A_R(p)\) &
\(A_S(p)\) &
\(A_G(p)\) &
\(A_C(p)\) &
\(PL_{A,\ell}(p)\) \\
\hline
1
& \(H\) & 2.653 & 0.237 & 0.551 & 0.453 & 3.666 \\
& \(G\) & 2.547 & 0.204 & 0.478 & 0.349 & 3.381 \\
\hline
0.1
& \(H\) & 10.138 & 0.383 & 0.607 & 0.752 & 11.504 \\
& \(G\) & 9.767 & 0.330 & 0.525 & 0.559 & 10.856 \\
\hline
0.01
& \(H\) & 27.300 & 0.569 & 0.647 & 0.982 & 28.935 \\
& \(G\) & 26.392 & 0.490 & 0.559 & 0.720 & 27.675 \\
\hline
\end{tabular}
\end{table}


\subsection{Results}
Fig.~\ref{fig:q_pdf_validation} validates the SPA-based PDF of the effective small-scale channel power \(Q\) given in~\eqref{eq:pdf_Q_spa}. Fig.~\ref{fig:q_pdf_validation}(a) corresponds to random RIS phase assignment, whereas Figs.~\ref{fig:q_pdf_validation}(b) and \ref{fig:q_pdf_validation}(c) show the LoS-aligned phase configuration for \(30\times30\) and \(40\times40\) RISs, respectively. Since the absolute power levels differ considerably among the considered configurations, the PDFs are normalized for visualization purposes. In all cases, the analytical results closely match the Monte Carlo simulations, confirming the accuracy of the proposed SPA-based distribution. Under random phase assignment, increasing the RIS size does not necessarily provide a coherent power gain, since the individual reflected components may combine destructively. Consequently, the distribution corresponding to the larger RIS may not exhibit an improvement over that of the smaller RIS. In contrast, under LoS-aligned phase configuration, the \(40\times40\) RIS yields a considerably larger average effective channel power than the \(30\times30\) RIS.

An additional observation is that the LoS-aligned configuration does not necessarily require fully instantaneous element-wise channel knowledge in the considered static geometry. The transmitter, receiver, and HAPS locations can be mapped to the corresponding propagation directions through the ECEF/ENU geometry, from which the relevant AoA/AoD parameters and steering vectors are obtained. Hence, the transmit beamformer and RIS phase configuration can be interpreted as geometry-aided or location-assisted designs based primarily on the deterministic LoS structure. This suggests that, in slowly varying HAPS-assisted deployments with reliable positioning information, a substantial portion of the coherent beamforming gain may be obtained from location-derived geometric information.

\begin{figure}[t!]
    \centering
    \includegraphics[width=3.5in]{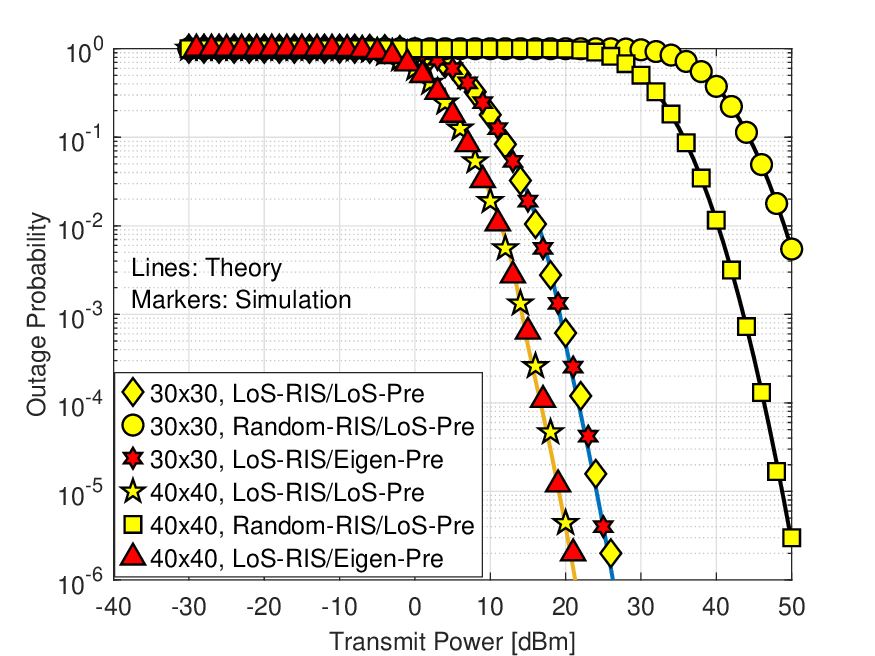}
    \caption{Outage probability performance of the HAPS-RIS-assisted multiple-antenna system for different RIS sizes under random, LoS-aligned, and eigenmode transmission at \(p=1\%\)~\((b=2)\).}
    \label{fig:HAPS_OP_EIG_LoS}
\end{figure}

Fig.~\ref{fig:HAPS_OP_EIG_LoS} further demonstrates the effectiveness of the proposed geometry-aided transmission strategy in terms of outage probability. For the eigenmode benchmark, the RIS phase configuration is kept LoS-aligned, while the transmit precoder is selected as the dominant right singular vector of the instantaneous cascaded channel. This benchmark therefore requires full instantaneous CSI together with an eigenvalue decomposition or singular value decomposition at each channel realization. In contrast, the LoS-RIS/LoS-Pre configuration constructs both the RIS phases and the transmit precoder from the deterministic LoS geometry. The outage curves of the LoS-RIS/LoS-Pre and LoS-RIS/Eigen-Pre configurations are nearly indistinguishable, showing that location-derived LoS information can provide transmit-precoding performance close to the instantaneous eigenmode benchmark in the considered HAPS-assisted scenario. The substantial performance gap between the LoS-RIS and Random-RIS cases further confirms the importance of coherent RIS phase alignment.

\begin{figure}[t!]
    \centering
    \includegraphics[width=3.5in]{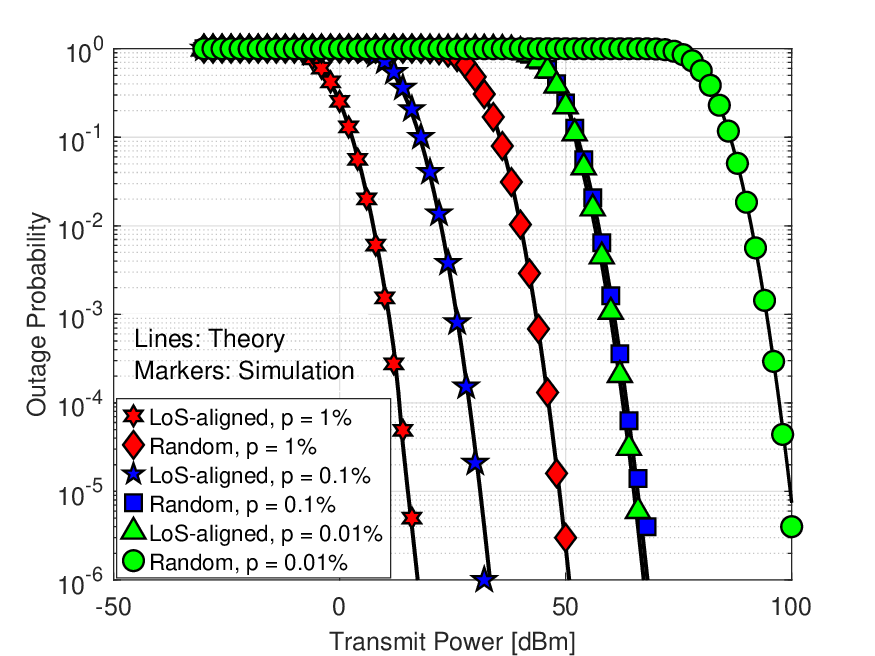}
    \caption{Outage probability of the \(50\times50\) HAPS-RIS-assisted system under different atmospheric exceedance probabilities \(p\) for random and LoS-aligned RIS configurations~\((b=2)\).}
    \label{fig:outage_atmospheric}
\end{figure}

Fig.~\ref{fig:outage_atmospheric} illustrates the impact of atmospheric attenuation on the outage performance of the \(50\times50\) HAPS-RIS-assisted system. As the exceedance probability decreases from \(p=1\%\) to \(p=0.01\%\), the corresponding increase in atmospheric attenuation shifts the outage curves toward substantially higher transmit-power levels. With the same LoS-oriented transmit precoder, the LoS-aligned RIS configuration consistently provides a pronounced power advantage over random phase assignment under all considered atmospheric conditions. This advantage remains substantial even under the most severe attenuation case, demonstrating that coherent exploitation of the deterministic LoS geometry can effectively compensate for a considerable portion of the additional propagation loss.

\begin{figure}[t!]
    \centering
    \includegraphics[width=3.5in]{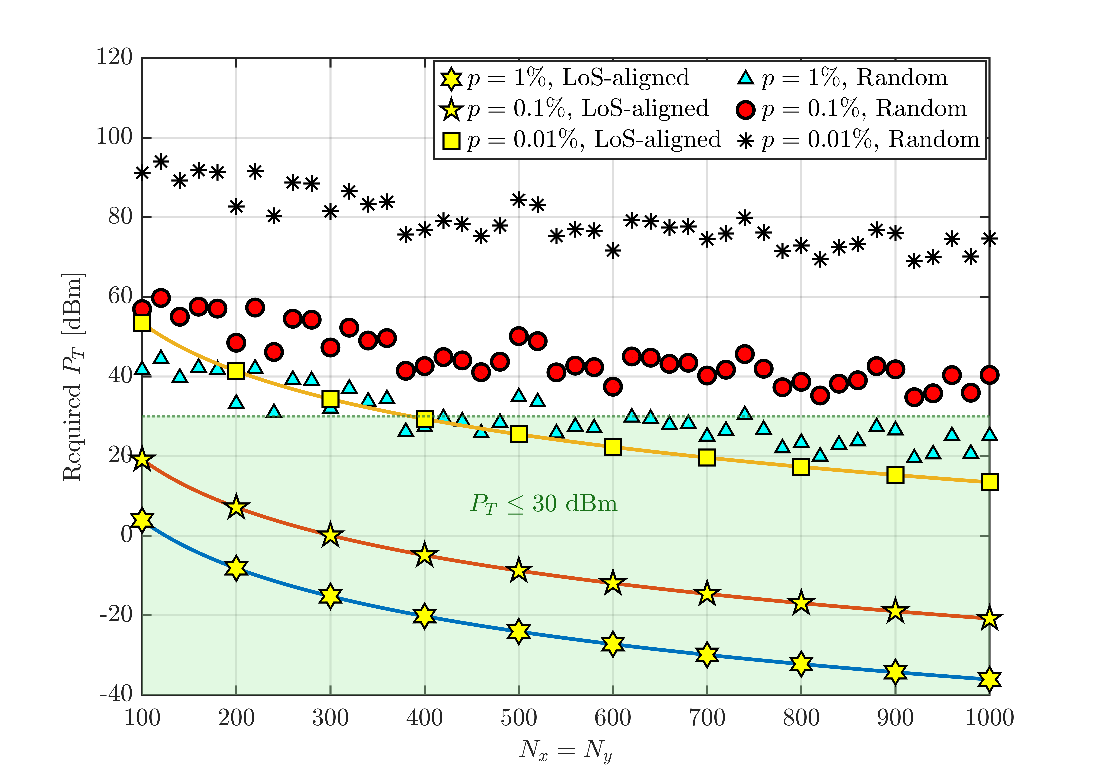}
    \caption{Required transmit power to achieve \(P_{\mathrm{out}}=10^{-5}\) as a function of RIS size and atmospheric exceedance probability \(p\) for LoS-aligned and random RIS phase configurations~\((b=2)\).}
    \label{fig:required_power_contour}
\end{figure}

The preceding figures employed moderate RIS dimensions to enable computationally feasible Monte Carlo simulations. Having established the close agreement between theory and simulation, Fig.~\ref{fig:required_power_contour} extends the analysis to a broader HAPS-RIS design range with \(N_x=N_y\in[100,1000]\). With a unit RIS element size of \(d_{\rm RIS}=\lambda_c/2\) and an inter-element spacing of \(\Delta_{\rm RIS}=5\lambda_c/8\), this corresponds approximately to physical side lengths from \(0.60\) m to \(6.04\) m at \(f_c=31\) GHz, thereby covering larger and more representative RIS apertures for HAPS-mounted deployments.

Fig.~\ref{fig:required_power_contour} presents the transmit power required to satisfy \(P_{\mathrm{out}}=10^{-5}\) as a function of RIS size and atmospheric exceedance probability \(p\). A clear contrast is observed between the LoS-aligned and random RIS configurations. Under random phase assignment, increasing the RIS size does not yield a systematic reduction in the required transmit power, indicating that aperture scaling alone is insufficient without coherent combining. In contrast, the LoS-aligned configuration provides substantial power savings, particularly for large RIS apertures and moderate atmospheric attenuation. Under favorable atmospheric conditions, the target reliability can be achieved at comparatively low transmit-power levels, whereas increasingly severe attenuation requires progressively larger apertures to recover the additional propagation loss. These results reveal a clear tradeoff among RIS aperture, atmospheric conditions, and transmit power, and emphasize the importance of coherent phase alignment for reliable HAPS-assisted links.

\begin{figure*}[t]
    \centering

    \begin{minipage}[c]{0.32\textwidth}
        \centering
        \includegraphics[width=\linewidth]{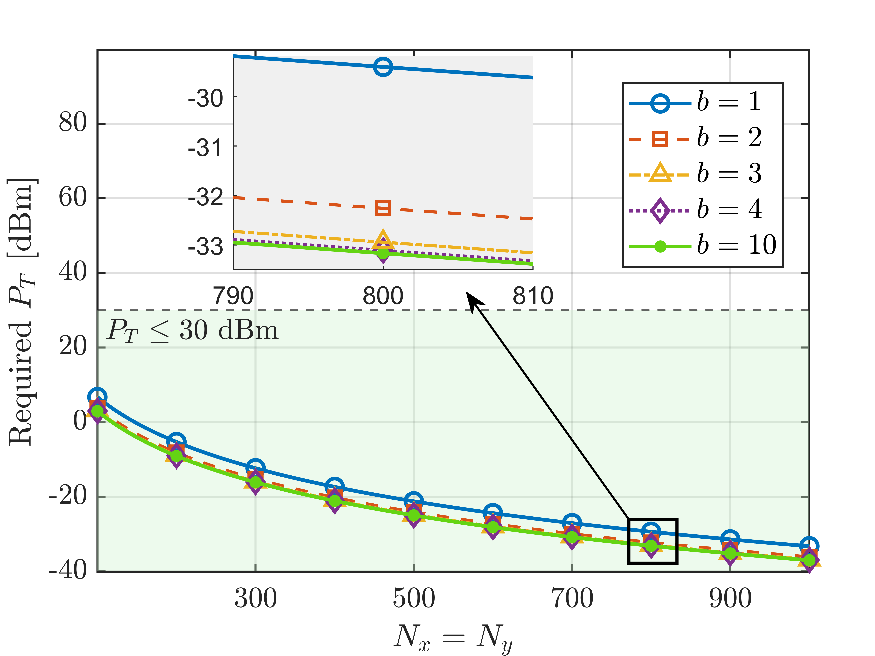}

        \small (a) \(p=1\%\)
    \end{minipage}
    \hfill
    \begin{minipage}[c]{0.32\textwidth}
        \centering
        \includegraphics[width=\linewidth]{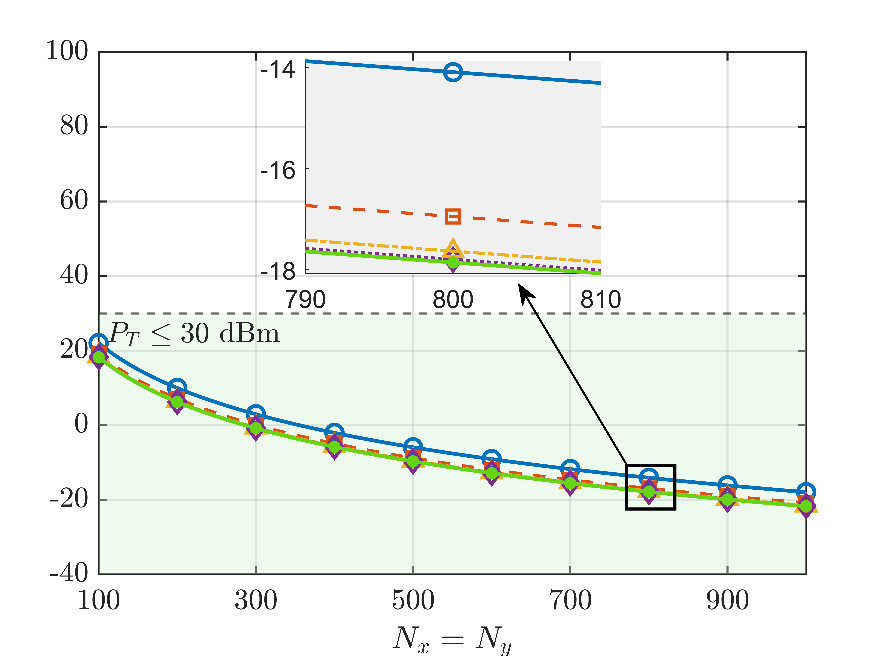}

        \small (b) \(p=0.1\%\)
    \end{minipage}
    \hfill
    \begin{minipage}[c]{0.32\textwidth}
        \centering
        \includegraphics[width=\linewidth]{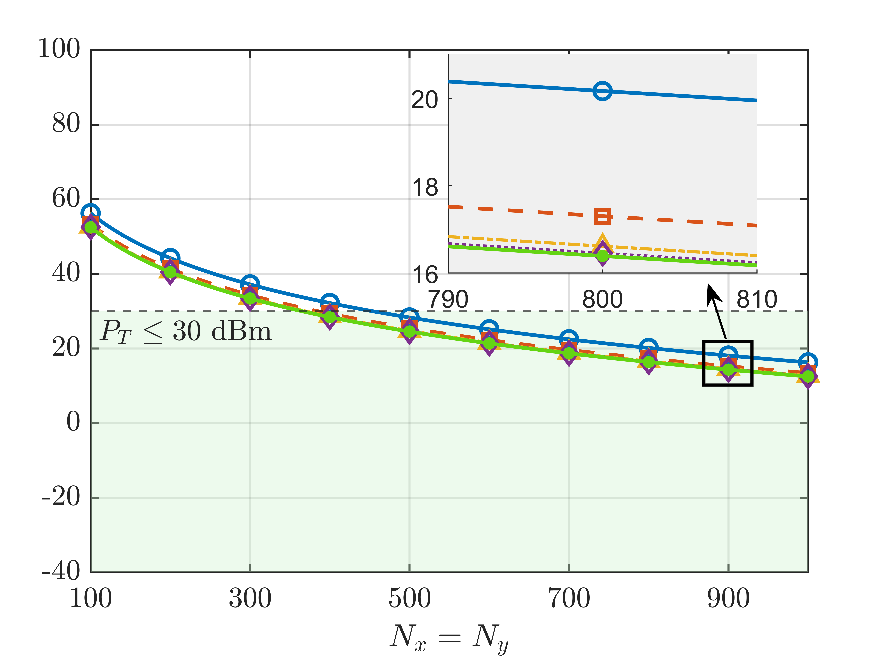}

        \small (c) \(p=0.01\%\)
    \end{minipage}

    \caption{Required transmit power to achieve \(P_{\mathrm{out}}=10^{-5}\) as a function of RIS size for different phase resolutions under LoS-aligned transmission.}
    \label{fig:required_power_phase_res}
\end{figure*}

Finally, Fig.~\ref{fig:required_power_phase_res} shows the impact of RIS phase resolution on the required transmit power under different atmospheric conditions. Increasing the phase resolution reduces the required \(P_T\), while the additional gain becomes negligible beyond approximately \(b=4\) bits. For \(N_x=N_y=800\), increasing the resolution from \(b=1\) to \(b=4\) yields approximately \(3.77\), \(3.70\), and \(3.76\) dB power savings for \(p=1\%\), \(0.1\%\), and \(0.01\%\), respectively. Hence, a gain of about \(3.8\) dB can be achieved across the considered atmospheric conditions, while further increasing the phase resolution provides only marginal improvement, highlighting a clear tradeoff between RIS hardware complexity and power efficiency.

\section{Conclusion}
\label{Sec:conclusions}

This paper developed a tractable performance-analysis framework for HAPS-RIS-assisted multiple-antenna systems over cascaded Rician--Rician channels with practical RIS impairments and location-dependent atmospheric attenuation. The effective small-scale channel power was characterized using SPA, while the composite log-normal large-scale effects were incorporated through Gauss--Hermite quadrature. The resulting analytical PDF, CDF, and outage expressions were shown to closely match Monte Carlo simulations, confirming the accuracy of the proposed framework under different RIS configurations and atmospheric conditions.

The numerical results also provided several practical design insights. By exploiting the deployment geometry through ECEF/ENU-based direction extraction, LoS-oriented transmit precoding and RIS phase alignment can be implemented using positioning information rather than full instantaneous CSI. In the considered scenarios, the resulting LoS-aligned precoder achieved outage performance nearly indistinguishable from dominant-eigenmode precoding. The results further showed that coherent RIS phase alignment is essential for reducing the required transmit power, whereas increasing the RIS size alone is insufficient under random phase assignment. Increasing the RIS phase resolution also reduces the required transmit power, although the additional gain becomes marginal beyond \(4\) bits, highlighting a practical balance between phase-shifter complexity and power efficiency. Under severe atmospheric attenuation, larger RIS apertures become necessary even with proper phase alignment, revealing a clear tradeoff between transmit power, RIS size, and propagation conditions. Although the numerical study considered a representative deployment in Türkiye, the proposed framework can be directly extended to other locations, HAPS altitudes, antenna configurations, and atmospheric conditions by recomputing the corresponding geometric and propagation parameters.

\appendices
\section{Derivation of the Mean and Variance in Lemma~\ref{lem:CLT_Rician_Heff}}
\label{app:CLT_derivation}

For the $(r,t)$-th entry of the cascaded channel in
Lemma~\ref{lem:CLT_Rician_Heff}, we write
\begin{equation}
\label{eq:app_frt}
\widehat f_{r,t}
=
\sum_{n=1}^{N_{\mathrm{RIS}}}
\widehat g_{r,n}\Gamma_n\widehat h_{n,t}.
\end{equation}
The Rician channel coefficients are decomposed as
\(
\widehat g_{r,n}
=
\bar g_{r,n}
+
\widetilde g_{r,n}
\) and \( \widehat h_{n,t}
=
\bar h_{n,t}
+
\widetilde h_{n,t},
\)
where $\bar g_{r,n}$ and $\bar h_{n,t}$ denote the deterministic LoS
components, while $\widetilde g_{r,n}$ and $\widetilde h_{n,t}$ denote
the zero-mean NLoS components. From the Rician channel models in
\eqref{eq:rician_H} and \eqref{eq:rician_G},
\(
\mathbb E[\widetilde g_{r,n}]
=
\mathbb E[\widetilde h_{n,t}]
=
0,
\)
with
\begin{equation}
\mathbb E
\left[
|\widetilde g_{r,n}|^2
\right]
=
\frac{\sigma_g^2}{K_G+1},
\qquad
\mathbb E
\left[
|\widetilde h_{n,t}|^2
\right]
=
\frac{\sigma_h^2}{K_H+1}.
\end{equation}

\subsubsection{Mean}Since the two hops are mutually independent and $\Gamma_n$ is
deterministic for a given RIS configuration, the mean of
\eqref{eq:app_frt} is
\begin{equation}
\mathbb E[\widehat f_{r,t}]=
\sum_{n=1}^{N_{\mathrm{RIS}}}
\Gamma_n
\mathbb E[\widehat g_{r,n}]
\mathbb E[\widehat h_{n,t}]
=
\sum_{n=1}^{N_{\mathrm{RIS}}}
\Gamma_n
\bar g_{r,n}
\bar h_{n,t}.
\label{eq:app_mean_sum}
\end{equation}
Using the rank-one LoS channel structures in
\eqref{eq:H_LoS} and \eqref{eq:G_LoS}, the deterministic components are
given by
\begin{subequations}
\begin{align}
&\bar g_{r,n}
=
\sqrt{\frac{K_G}{K_G+1}}
\big[\mathbf a_{\mathrm R}(\vartheta_r)\big]_r
\big[\mathbf a_{\mathrm{RIS}}(\phi_G,\theta_G)\big]_n^{*},
\\
&
\bar h_{n,t}
=
\sqrt{\frac{K_H}{K_H+1}}
\big[\mathbf a_{\mathrm{RIS}}(\phi_H,\theta_H)\big]_n
\big[\mathbf a_{\mathrm T}(\phi_t,\theta_t)\big]_t^{*}.
\end{align}
\end{subequations}
Substituting these expressions into \eqref{eq:app_mean_sum} yields
\begin{align}
\widehat{\mu}_{r,t}
&=
\sqrt{
\frac{K_HK_G}
{(K_H+1)(K_G+1)}
}
\big[\mathbf a_{\mathrm R}(\vartheta_r)\big]_r
\big[\mathbf a_{\mathrm T}(\phi_t,\theta_t)\big]_t^{*}
\nonumber\\
&\quad\times
\mathbf a_{\mathrm{RIS}}^{H}(\phi_G,\theta_G)
\boldsymbol{\Phi}
\mathbf a_{\mathrm{RIS}}(\phi_H,\theta_H)
\nonumber\\
&=
\sqrt{
\frac{K_HK_G}
{(K_H+1)(K_G+1)}
}
\eta_{\mathrm{RIS}}
\big[\mathbf a_{\mathrm R}(\vartheta_r)\big]_r
\big[\mathbf a_{\mathrm T}(\phi_t,\theta_t)\big]_t^{*},
\label{eq:app_mean_final}
\end{align}
which gives the mean stated in Lemma~\ref{lem:CLT_Rician_Heff}.

\subsubsection{Variance}To derive the variance, the second moment of
$\widehat f_{r,t}$ is first written as
\begin{align}
\mathbb E
\left[
|\widehat f_{r,t}|^2
\right]
&=
\sum_{n=1}^{N_{\mathrm{RIS}}}
|\Gamma_n|^2
\mathbb E
\left[
|\widehat g_{r,n}|^2
\right]
\mathbb E
\left[
|\widehat h_{n,t}|^2
\right]
\nonumber\\
&\quad+
\sum_{\substack{n,m=1\\n\neq m}}^{N_{\mathrm{RIS}}}
\Gamma_n\Gamma_m^{*}
\mathbb E
\left[
\widehat g_{r,n}
\widehat g_{r,m}^{*}
\right]
\mathbb E
\left[
\widehat h_{n,t}
\widehat h_{m,t}^{*}
\right].
\label{eq:app_second_moment}
\end{align}
Since the NLoS components are independent across the RIS index and have
zero mean, for $n\neq m$,
\begin{subequations}
\begin{align}
&\mathbb E
\left[
\widehat g_{r,n}
\widehat g_{r,m}^{*}
\right]
=
\bar g_{r,n}\bar g_{r,m}^{*},
\\&
\mathbb E
\left[
\widehat h_{n,t}
\widehat h_{m,t}^{*}
\right]
=
\bar h_{n,t}\bar h_{m,t}^{*}.
\end{align}
\end{subequations}
Hence, the cross-term contribution in \eqref{eq:app_second_moment}
satisfies
\begin{align}
&\sum_{\substack{n,m=1\\n\neq m}}^{N_{\mathrm{RIS}}}
\Gamma_n\Gamma_m^{*}
\bar g_{r,n}\bar g_{r,m}^{*}
\bar h_{n,t}\bar h_{m,t}^{*}
\nonumber\\
&\qquad=
|\widehat{\mu}_{r,t}|^2
-
\sum_{n=1}^{N_{\mathrm{RIS}}}
|\Gamma_n|^2
|\bar g_{r,n}|^2
|\bar h_{n,t}|^2.
\label{eq:app_cross_terms}
\end{align}
Therefore,
\begin{align}
\widehat{\Sigma}_{r,t}
&=
\mathrm{Var}(\widehat f_{r,t})
=
\mathbb E
\left[
|\widehat f_{r,t}|^2
\right]
-
|\widehat{\mu}_{r,t}|^2
\nonumber\\
&=
\sum_{n=1}^{N_{\mathrm{RIS}}}
|\Gamma_n|^2
\Big(
\mathbb E[|\widehat g_{r,n}|^2]
\mathbb E[|\widehat h_{n,t}|^2]
-
|\bar g_{r,n}|^2
|\bar h_{n,t}|^2
\Big).
\label{eq:app_variance_identity}
\end{align}
Using
\begin{subequations}
\begin{align}
&\mathbb E
\left[
|\widehat g_{r,n}|^2
\right]
=
|\bar g_{r,n}|^2
+
\frac{\sigma_g^2}{K_G+1},
\\
&\mathbb E
\left[
|\widehat h_{n,t}|^2
\right]
=
|\bar h_{n,t}|^2
+
\frac{\sigma_h^2}{K_H+1},
\end{align}
\end{subequations}
together with $|\Gamma_n|^2=\beta_n^2$, \eqref{eq:app_variance_identity}
becomes
\begin{align}
\widehat{\Sigma}_{r,t}
=
\sum_{n=1}^{N_{\mathrm{RIS}}}
\beta_n^2
\Bigg(
&
\frac{\sigma_g^2\sigma_h^2}
{(K_G+1)(K_H+1)}
+
\frac{
|\bar g_{r,n}|^2\sigma_h^2
}{
K_H+1
}
\nonumber\\
&+
\frac{
|\bar h_{n,t}|^2\sigma_g^2
}{
K_G+1
}
\Bigg).
\label{eq:app_variance_general}
\end{align}
Since the adopted array steering vectors have unit-modulus entries,
\(
|\bar g_{r,n}|^2
=
\frac{K_G}{K_G+1},\)
\(
|\bar h_{n,t}|^2
=
\frac{K_H}{K_H+1}.
\)
For unit-variance NLoS components, i.e.,
$\sigma_g^2=\sigma_h^2=1$, \eqref{eq:app_variance_general} reduces to
\begin{equation}
\widehat{\Sigma}_{r,t}
=
\frac{K_H+K_G+1}
{(K_H+1)(K_G+1)}
\sum_{n=1}^{N_{\mathrm{RIS}}}
\beta_n^2.
\label{eq:app_variance_final}
\end{equation}
Therefore, the variance is independent of the transmit and receive antenna
indices $(r,t)$. Since \eqref{eq:app_frt} is a sum of independent
contributions across the RIS elements, the CLT applies for a
sufficiently large $N_{\mathrm{RIS}}$. Together with the derived mean
and variance, this establishes the Gaussian approximation stated in
Lemma~\ref{lem:CLT_Rician_Heff}.
\hfill$\blacksquare$

\section{Derivation of the Covariance Matrix~\eqref{eq:Reff_def}}
\label{app:Reff_derivation}

This appendix derives the structured covariance matrix in
\eqref{eq:Reff_def}. To simplify the notation, we define
\(
\mathbf a_H \triangleq \mathbf a_{\mathrm{RIS}}(\phi_H,\theta_H),~
\mathbf a_G \triangleq \mathbf a_{\mathrm{RIS}}(\phi_G,\theta_G).
\)
Using the Rician channel models in \eqref{eq:rician_H} and
\eqref{eq:rician_G}, the small-scale channels can be written as
\(
\widehat{\mathbf H}
=
\overline{\mathbf H}
+
\widetilde{\mathbf H}
\) and \(\widehat{\mathbf G}
=
\overline{\mathbf G}
+
\widetilde{\mathbf G},
\)
where
\begin{align}
\overline{\mathbf H}
&=
\sqrt{\frac{K_H}{K_H+1}}\mathbf a_H\mathbf m_t^H,
&
\widetilde{\mathbf H}
&=
\sqrt{\frac{1}{K_H+1}}\mathbf H_{\mathrm{NLoS}},
\\
\overline{\mathbf G}
&=
\sqrt{\frac{K_G}{K_G+1}}\mathbf m_r\mathbf a_G^H,
&
\widetilde{\mathbf G}
&=
\sqrt{\frac{1}{K_G+1}}\mathbf G_{\mathrm{NLoS}}.
\end{align}
Accordingly, the entries of $\widetilde{\mathbf H}$ and
$\widetilde{\mathbf G}$ are zero-mean complex Gaussian random variables with
variances $\sigma_h^2/(K_H+1)$ and $\sigma_g^2/(K_G+1)$, respectively.
Under the LoS-aligned beamformer in \eqref{eq:precoder_mt}, the projected transmitter--RIS channel can be expressed as
\(
\widehat{\mathbf h}
=
\widehat{\mathbf H}\mathbf w
=
\overline{\mathbf h}
+
\widetilde{\mathbf h},
\)
where
\begin{subequations}
\begin{align}
&\overline{\mathbf h}
=
\overline{\mathbf H}\mathbf w
=
\sqrt{\frac{K_H}{K_H+1}}
\|\mathbf m_t\|\mathbf a_H,
\\
&\widetilde{\mathbf h}
=
\widetilde{\mathbf H}\mathbf w
\sim
\mathcal{CN}
\left(
\mathbf 0,
\frac{\sigma_h^2}{K_H+1}\mathbf I_{N_{\mathrm{RIS}}}
\right).
\end{align}
\end{subequations}
The effective small-scale channel vector after RIS reflection and
LoS-aligned precoding is given by
\(
\widehat{\mathbf h}_{\mathrm{eff}}
=
\widehat{\mathbf F}\mathbf w
=
\widehat{\mathbf G}\boldsymbol{\Phi}\widehat{\mathbf H}\mathbf w.
\)
Substituting the LoS and NLoS components yields
\begin{equation}
\widehat{\mathbf F}\mathbf w
=
\overline{\mathbf G}\boldsymbol{\Phi}\overline{\mathbf h}
+
\overline{\mathbf G}\boldsymbol{\Phi}\widetilde{\mathbf h}
+
\widetilde{\mathbf G}\boldsymbol{\Phi}\overline{\mathbf h}
+
\widetilde{\mathbf G}\boldsymbol{\Phi}\widetilde{\mathbf h}.
\end{equation}
The first term is deterministic and corresponds to the mean vector
\(
\boldsymbol{\mu}
=
\mathbb E[\widehat{\mathbf F}\mathbf w]
=
\overline{\mathbf G}\boldsymbol{\Phi}\overline{\mathbf h}.
\)
This expression coincides with the deterministic LoS mean component
$\boldsymbol{\mu}$ introduced in \eqref{eq:z_and_mu}.
Thus, the zero-mean fluctuation term is written as
\begin{equation}
\mathbf z
=
\widehat{\mathbf F}\mathbf w-\boldsymbol{\mu}
=
\mathbf z_1+\mathbf z_2+\mathbf z_3,
\end{equation}
where
\(
\mathbf z_1
=
\overline{\mathbf G}\boldsymbol{\Phi}\widetilde{\mathbf h},\)
\(\mathbf z_2
=
\widetilde{\mathbf G}\boldsymbol{\Phi}\overline{\mathbf h},\) and
\(\mathbf z_3
=
\widetilde{\mathbf G}\boldsymbol{\Phi}\widetilde{\mathbf h}.
\)
Since the NLoS components of the two hops are independent and zero-mean, the
cross-covariance terms vanish. Therefore,
\begin{equation}
\mathbf R_{\mathrm{eff}}
=
\mathbb E[\mathbf z\mathbf z^H]
=
\sum_{i=1}^{3}
\mathbb E[\mathbf z_i\mathbf z_i^H].
\end{equation}

For $\mathbf z_1$, using
$\mathbb E[\widetilde{\mathbf h}\widetilde{\mathbf h}^H]
=
\frac{\sigma_h^2}{K_H+1}\mathbf I_{N_{\mathrm{RIS}}}$, we obtain
\begin{align}
\mathbb E[\mathbf z_1\mathbf z_1^H]
&=
\overline{\mathbf G}\boldsymbol{\Phi}
\mathbb E[\widetilde{\mathbf h}\widetilde{\mathbf h}^H]
\boldsymbol{\Phi}^H\overline{\mathbf G}^H
\nonumber\\
&=
\frac{\sigma_h^2}{K_H+1}
\overline{\mathbf G}
\boldsymbol{\Phi}\boldsymbol{\Phi}^H
\overline{\mathbf G}^H.
\end{align}
Since
\(
\boldsymbol{\Phi}\boldsymbol{\Phi}^H
=
\mathrm{diag}(\beta_1^2,\ldots,\beta_{N_{\mathrm{RIS}}}^2),
\)
and the RIS steering vectors have unit-modulus entries, using
\(\xi_{\beta}\) defined in \eqref{eq:xi_beta} yields
\begin{equation}
\mathbb E[\mathbf z_1\mathbf z_1^H]
=
\frac{\sigma_h^2K_G}{(K_H+1)(K_G+1)}
\xi_{\beta}
\mathbf m_r\mathbf m_r^H.
\end{equation}

For $\mathbf z_2$, we use the identity~\cite[Theorem~3.2b.1]{Quadratic_Forms}
\begin{equation}
\mathbb E[
\widetilde{\mathbf G}\mathbf A\widetilde{\mathbf G}^H]
=
\frac{\sigma_g^2}{K_G+1}
\mathrm{tr}(\mathbf A)\mathbf I_{N_r},
\end{equation}
where
\(
\mathbf A
=
\boldsymbol{\Phi}
\overline{\mathbf h}\overline{\mathbf h}^H
\boldsymbol{\Phi}^H.
\)
Then,
\begin{align}
\mathbb E[\mathbf z_2\mathbf z_2^H]
&=
\frac{\sigma_g^2}{K_G+1}
\mathrm{tr}
\left(
\boldsymbol{\Phi}
\overline{\mathbf h}\overline{\mathbf h}^H
\boldsymbol{\Phi}^H
\right)
\mathbf I_{N_r}
\nonumber\\
&=
\frac{\sigma_g^2K_H\|\mathbf m_t\|^2}
{(K_G+1)(K_H+1)}
\xi_{\beta}
\mathbf I_{N_r}.
\end{align}

For $\mathbf z_3$, conditioning on $\widetilde{\mathbf h}$ and applying the
same trace identity gives
\begin{align}
\mathbb E[\mathbf z_3\mathbf z_3^H]
&=
\mathbb E_{\widetilde{\mathbf h}}
\left[
\mathbb E_{\widetilde{\mathbf G}}
\left[
\widetilde{\mathbf G}
\boldsymbol{\Phi}
\widetilde{\mathbf h}\widetilde{\mathbf h}^H
\boldsymbol{\Phi}^H
\widetilde{\mathbf G}^H
\right]
\right]
\nonumber\\
&=
\frac{\sigma_g^2}{K_G+1}
\mathbb E_{\widetilde{\mathbf h}}
\left[
\mathrm{tr}
\left(
\boldsymbol{\Phi}
\widetilde{\mathbf h}\widetilde{\mathbf h}^H
\boldsymbol{\Phi}^H
\right)
\right]
\mathbf I_{N_r}
\nonumber\\
&=
\frac{\sigma_g^2}{K_G+1}
\mathrm{tr}
\left(
\boldsymbol{\Phi}
\mathbb E[
\widetilde{\mathbf h}\widetilde{\mathbf h}^H]
\boldsymbol{\Phi}^H
\right)
\mathbf I_{N_r}
\nonumber\\
&=
\frac{\sigma_g^2}{K_G+1}
\mathrm{tr}
\left(
\boldsymbol{\Phi}
\frac{\sigma_h^2}{K_H+1}\mathbf I_{N_{\mathrm{RIS}}}
\boldsymbol{\Phi}^H
\right)
\mathbf I_{N_r}
\nonumber\\
&=
\frac{\sigma_g^2\sigma_h^2}
{(K_G+1)(K_H+1)}
\xi_{\beta}
\mathbf I_{N_r}.
\end{align}
Combining the three covariance contributions yields the structured covariance matrix in \eqref{eq:Reff_def}.
\hfill$\blacksquare$

\section{Derivation of PDF and CDF Approximations \eqref{eq:snr_pdf_GH} and \eqref{eq:snr_cdf_GH}}
\label{app:GH_outage}

Using the log-normal representation of \(L_{\mathrm{tot}}\) in
\eqref{eq:Ltot_lognormal}, the
unconditional received SNR PDF in \eqref{eq:snr_pdf_integral} can be
written as
\begin{align}
\label{eq:app_snr_pdf_gaussian}
f_{\rho}(\rho)
&=
\frac{1}{\sqrt{2\pi}}
\int_{-\infty}^{\infty}
\frac{
\sigma^2
}{
P_T
\exp\left(
\mu_L(p)+\sigma_Lx
\right)
}
\nonumber\\
&\quad\times
f_Q\left(
\frac{
\rho\sigma^2
}{
P_T
\exp\left(
\mu_L(p)+\sigma_Lx
\right)
}
\right)
\exp\left(\frac{-x^2}{2}\right)
\,dx.
\end{align}
Similarly, the corresponding SNR CDF becomes
\begin{align}
\label{eq:app_snr_cdf_gaussian}
F_{\rho}(\rho)
=
\frac{1}{\sqrt{2\pi}}
\int_{-\infty}^{\infty}
&F_Q\left(
\frac{
\rho\sigma^2
}{
P_T
\exp\left(
\mu_L(p)+\sigma_Lx
\right)
}
\right)
\nonumber \\ 
&\times\exp\left(\frac{-x^2}{2}\right)
\,dx.
\end{align}
Applying the change of variables \(x=\sqrt{2}t\) to
\eqref{eq:app_snr_pdf_gaussian}, the PDF expression becomes
\begin{align}
\label{eq:app_snr_pdf_hermite}
f_{\rho}(\rho)
&=
\frac{1}{\sqrt{\pi}}
\int_{-\infty}^{\infty}
\frac{
\sigma^2
}{
P_T
\exp\left(
\mu_L(p)+\sqrt{2}\sigma_Lt
\right)
}
\nonumber\\
&\quad\times
f_Q\left(
\frac{
\rho\sigma^2
}{
P_T
\exp\left(
\mu_L(p)+\sqrt{2}\sigma_Lt
\right)
}
\right)
\exp\left(-t^2\right)
\,dt.
\end{align}
Likewise, \eqref{eq:app_snr_cdf_gaussian} becomes
\begin{align}
\label{eq:app_snr_cdf_hermite}
F_{\rho}(\rho)
=
\frac{1}{\sqrt{\pi}}
\int_{-\infty}^{\infty}
&F_Q\left(
\frac{
\rho\sigma^2
}{
P_T
\exp\left(
\mu_L(p)+\sqrt{2}\sigma_Lt
\right)
}
\right) \nonumber \\
& \times \exp\left(-t^2\right)
\,dt.
\end{align}

The integrals in \eqref{eq:app_snr_pdf_hermite} and
\eqref{eq:app_snr_cdf_hermite} are now in the standard
Gauss--Hermite form. Let \(x_i\) denote the \(i\)th abscissa of the
\(N_{\mathrm{GH}}\)-point Gauss--Hermite quadrature rule, corresponding
to the \(i\)th root of the Hermite polynomial
\(H_{N_{\mathrm{GH}}}(x)\)
\cite[eqs.~(22.1.2) and (22.2.14)]{Abramowitz}. The corresponding
weight is
\begin{equation}
\label{eq:app_GH_weight}
w_i
=
\frac{
2^{N_{\mathrm{GH}}-1}N_{\mathrm{GH}}!\sqrt{\pi}
}{
N_{\mathrm{GH}}^2
\left[
H_{N_{\mathrm{GH}}-1}(x_i)
\right]^2
}.
\end{equation}
Using the Gauss--Hermite quadrature rule in \cite[eq.~(25.4.46)]{Abramowitz}, the integrals in \eqref{eq:app_snr_pdf_hermite} and \eqref{eq:app_snr_cdf_hermite} yield the SNR PDF and CDF approximations in \eqref{eq:snr_pdf_GH} and \eqref{eq:snr_cdf_GH}, respectively. This completes the derivation.
\hfill$\blacksquare$

\bibliographystyle{IEEEtran}
\bibliography{ref}

@article{RIS-AMP-RES,
  author = {Abeywickrama, S. and Zhang, R. and Wu, Q. and Yuen, C.},
  title = {Intelligent Reflecting Surface: Practical Phase Shift Model and Beamforming Optimization},
  journal = {{IEEE} Trans. Commun.},
  volume = {68},
  number = {9},
  pages = {5849--5863},
  year = {2020}
}

@book{Abramowitz,
  author = {Abramowitz, M. and Stegun, I. A.},
  title = {Handbook of Mathematical Functions},
  publisher = {Dover},
  year = {1974}
}

@ARTICLE{HAPS_Connectivity,
  author={Khoshafa, Majid H. and \emph{et al.}},
  journal={{IEEE} Commun. Surveys Tuts.}, 
  title={{RIS}-Assisted Physical Layer Security in Emerging {RF} and Optical Wireless Communications Systems: A Comprehensive Survey}, 
  year={2025},
  volume={27},
  number={4},
  pages={2156-2203}}

@ARTICLE{G_Kurt_HAPS,
  author={Karabulut Kurt, Gunes and \emph{et al.}},
  journal={{IEEE} Commun. Surveys Tuts.}, 
  title={A Vision and Framework for the High Altitude Platform Station {(HAPS)} Networks of the Future}, 
  year={2021},
  volume={23},
  number={2},
  pages={729-779}}

@ARTICLE{HAPS_RIS_Intro,
  author={Alfattani, Safwan and Yadav, Animesh and Yanikomeroglu, Halim and Yongaçoglu, Abbas},
  journal={{IEEE} Wireless Commun. Lett.}, 
  title={Resource-Efficient {HAPS-RIS} Enabled Beyond-Cell Communications}, 
  year={2023},
  volume={12},
  number={4},
  pages={679-683}}

@ARTICLE{HAPS_RIS_Magazine_2,
  author={Ye, Jia and Qiao, Jingping and Kammoun, Abla and Alouini, Mohamed-Slim},
  journal={Proc. IEEE}, 
  title={Nonterrestrial Communications Assisted by Reconfigurable Intelligent Surfaces}, 
  year={2022},
  volume={110},
  number={9},
  pages={1423-1465}}

@ARTICLE{HAPS_Intro1,
  author = {Abbasi, Omid and Yadav, Animesh and Yanikomeroglu, Halim and Dao, Ngoc-Dung and Senarath, Gamini and Zhu, Peiying},
  journal={{IEEE} Wireless Commun.}, 
  title={{HAPS for 6G} Networks: {Potential} Use Cases, Open Challenges, and Possible Solutions}, 
  year={2024},
  volume={31},
  number={3},
  pages={324-331},
  doi={10.1109/MWC.012.2200365}}

@ARTICLE{HAPS_Intro2,
  author={Azizi, Arman and Farhang, Arman},
  journal={{IEEE} Wireless Commun. Lett.}, 
  title={{RIS} Meets Aerodynamic {HAPS: A} Multi-Objective Optimization Approach}, 
  year={2023},
  volume={12},
  number={11},
  pages={1851-1855},
  doi={10.1109/LWC.2023.3296023}}

@ARTICLE{1,
  author={Memarian, Hanieh and Mohammad Razavizadeh, S. and Kuhestani, Ali},
  journal={IEEE Access}, 
  title={Enhancing Physical Layer Security in {RIS}-Aided {HAPS} for Non-Terrestrial Networks}, 
  year={2025},
  volume={13},
  number={},
  pages={103010-103018},
  doi={10.1109/ACCESS.2025.3576866}}

@ARTICLE{2,
  author={Al-Rimawi, Ashraf and Al-Dweik, Arafat and Hamila, Ridha and Gouissem, Ala},
  journal={IEEE Trans. Aerosp. Electron. Syst.}, 
  title={Outage Probability Analysis of {RIS}-Assisted {UAV} Communications with Direct Link and {ISI}}, 
  year={2025},
  volume={},
  number={},
  pages={1-8},
  doi={10.1109/TAES.2025.3639189}}

@ARTICLE{3,
  author={Odeyemi, Kehinde Oluwasesan and Owolawi, Pius Adewale and Olakanmi, Oladayo Olufemi},
  journal={IEEE Access}, 
  title={Reconfigurable Intelligent Surface-Assisted HAPS Relaying Communication Networks for Multiusers Under {AF} Protocol: {A} Performance Analysis}, 
  year={2022},
  volume={10},
  number={},
  pages={14857-14869},
  doi={10.1109/ACCESS.2022.3146885}}

@ARTICLE{4,
  author={Abbasi Mosleh, Marjan and Heliot, Fabien and Tafazolli, Rahim},
  journal={{IEEE} Commun. Lett.}, 
  title={Ergodic Capacity Analysis of Reconfigurable Intelligent Surface Assisted {MIMO} Systems Over {Rayleigh-Rician} Channels}, 
  year={2023},
  volume={27},
  number={1},
  pages={75-79},
  doi={10.1109/LCOMM.2022.3221158}}

@ARTICLE{5,
  author={Kota, Kali Krishna and Mankar, Praful D. and Dhillon, Harpreet S.},
  journal={{IEEE} Wireless Commun. Lett.}, 
  title={Characterization of Capacity and Outage of {RIS}-Aided Downlink Systems Under {Rician} Fading}, 
  year={2025},
  volume={14},
  number={3},
  pages={631-635},
  doi={10.1109/LWC.2024.3518357}}

@ARTICLE{6,
  author={Alnajjar, Khawla A. and Abdelaziz Salem, A. and Ansari, Sam and El-Tarhuni, Mohamed},
  journal={IEEE Open J. Commun. Soc.}, 
  title={Enhanced Hybrid {HAP}-Assisted {NOMA-RIS} Systems for Next-Generation Wireless Networks}, 
  year={2025},
  volume={6},
  number={},
  pages={10694-10705},
  doi={10.1109/OJCOMS.2025.3643484}}

@ARTICLE{HAPS_RIS_Magazine_1,
  author={Karaman, Bilal and \emph{et al.}},
  journal={IEEE Internet Things Mag.}, 
  title={{HAPS-RIS}-Assisted {IoT} Networks for Disaster Recovery and Emergency Response: {Architecture}, Application Scenarios, and Open Challenges}, 
  year={2026},
  volume={},
  number={},
  pages={1-8},
  doi={10.1109/MIOT.2026.3673492}}

@ARTICLE{5G_Precoding_Survey,
  author={Ning, Boyu and \emph{et al.}},
  journal={IEEE Commun. Surveys and Tuts.}, 
  title={Precoding Matrix Indicator in the {5G NR} Protocol: {A} Tutorial on {3GPP} Beamforming Codebooks}, 
  year={2026},
  volume={},
  number={},
  pages={1-1},
  doi={10.1109/COMST.2026.3653568}}

@ARTICLE{HAPS_RIS_NOMA,
  author={Ji, Pingping and Jiang, Lingge and He, Chen and Lian, Zhuxian and He, Di},
  journal={IEEE Wireless Commun. Lett.}, 
  title={Active {RIS} Aided {NOMA} for {HAP-MISO} Systems}, 
  year={2024},
  volume={13},
  number={8},
  pages={2170-2174}}

@ARTICLE{Mosleh_2026,
  author={Mosleh, Marjan Abbasi and H´eliot, Fabien and Gradoni, Gabriele and Tafazolli, Rahim},
  journal={IEEE Trans. Veh. Technol.}, 
  title={Ergodic Capacity Analysis of {RIS}-Aided {MIMO} Systems under Amplitude-Phase Coupling}, 
  year={2026},
  volume={},
  number={},
  pages={1-16},
  doi={10.1109/TVT.2026.3684721}}

@ARTICLE{PL_RIS,
  author={Tang, Wankai and \emph{et al.}},
  journal={IEEE Trans. Wireless Commun.}, 
  title={Wireless Communications With Reconfigurable Intelligent Surface: {Path} Loss Modeling and Experimental Measurement}, 
  year={2021},
  volume={20},
  number={1},
  pages={421-439},
  doi={10.1109/TWC.2020.3024887}}

@article{daniels1954saddlepoint,
  title={Saddlepoint approximations in statistics},
  author={Daniels, Henry E},
  journal={The Annals of Mathematical Statistics},
  pages={631--650},
  year={1954},
  publisher={JSTOR}
}

@ARTICLE{abbasi2024ergodic,
  author={Mosleh, Marjan Abbasi and Héliot, Fabien and Tafazolli, Rahim},
  journal={IEEE Trans. Veh. Technol.}, 
  title={Ergodic Capacity Analysis of Reconfigurable Intelligent Surface Assisted {MIMO} Systems With a Practical Phase Shift and Amplitude Response}, 
  year={2024},
  volume={73},
  number={8},
  pages={11441-11457},
  doi={10.1109/TVT.2024.3376071}}

@book{Quadratic_Forms,
  author    = {Mathai, A. M. and Provost, S. B.},
  title     = {Quadratic Forms in Random Variables: Theory and Applications},
  series    = {Statistics: Textbooks and Monographs},
  volume    = {126},
  publisher = {Marcel Dekker, Inc.},
  address   = {New York, NY},
  year      = {1992}
}

@techreport{3GPP_TR_38811,
  author      = {{3rd Generation Partnership Project (3GPP)}},
  title       = {{Study on New Radio (NR) to Support Non-Terrestrial Networks}},
  institution = {{3GPP}},
  number      = {{TR 38.811 V15.1.0}},
  year        = {2019},
  month       = {Jun.}
}

@techreport{ITU_R_P618,
  author      = {{International Telecommunication Union}},
  title       = {{Propagation Data and Prediction Methods Required for the Design of Earth-Space Telecommunication Systems}},
  institution = {{ITU-R}},
  number      = {{Recommendation ITU-R P.618-14}},
  year        = {2023},
  month       = {Aug.}
}

@techreport{ITU_R_P676,
  author      = {{International Telecommunication Union}},
  title       = {{Attenuation by Atmospheric Gases and Related Effects}},
  institution = {{ITU-R}},
  number      = {{Recommendation ITU-R P.676-13}},
  year        = {2022},
  month       = {Aug.}
}

@techreport{ITU_R_P840,
  author      = {{International Telecommunication Union}},
  title       = {{Attenuation due to Clouds and Fog}},
  institution = {{ITU-R}},
  number      = {{Recommendation ITU-R P.840-9}},
  year        = {2023},
  month       = {Aug.},
  url         = {https://www.itu.int/rec/R-REC-P.840}
}

@techreport{ITU_R_P837,
  author      = {{ITU-R}},
  title       = {{Characteristics of Precipitation for Propagation Modelling}},
  institution = {International Telecommunication Union},
  number      = {Recommendation ITU-R P.837-8},
  year        = {2025},
  month       = sep,
  url = {https://www.itu.int/rec/R-REC-P}
}

@techreport{ITU_R_P838,
  author      = {{ITU-R}},
  title       = {{Specific Attenuation Model for Rain for Use in Prediction Methods}},
  institution = {International Telecommunication Union},
  number      = {Recommendation ITU-R P.838-3},
  year        = {2005},
  url = {https://www.itu.int/rec/R-REC-P}
}

@techreport{ITU_R_P839,
  author      = {{ITU-R}},
  title       = {{Rain Height Model for Prediction Methods}},
  institution = {International Telecommunication Union},
  number      = {Recommendation ITU-R P.839-4},
  year        = {2013},
  month       = sep,
  url = {https://www.itu.int/rec/R-REC-P}
}

@techreport{ITU_R_P453,
  author      = {{ITU-R}},
  title       = {{The Radio Refractive Index: Its Formula and Refractivity Data}},
  institution = {International Telecommunication Union},
  number      = {Recommendation ITU-R P.453},
  year        = {2023},
  url = {https://www.itu.int/rec/R-REC-P}
}

@techreport{ITU_R_P2145,
  author      = {{ITU-R}},
  title       = {{Digital Maps Related to the Calculation of Gaseous Attenuation and Related Effects}},
  institution = {International Telecommunication Union},
  number      = {Recommendation ITU-R P.2145-0},
  year        = {2022},
  month       = aug,
   url = {https://www.itu.int/rec/R-REC-P.2145}
}

@ARTICLE{AHO-13,
  author={Alfattani, Safwan and Jaafar, Wael and Hmamouche, Yassine and Yanikomeroglu, Halim and Yongaçoglu, Abbas},
  journal={IEEE Open J. Commun. Soc.}, 
  title={Link Budget Analysis for Reconfigurable Smart Surfaces in Aerial Platforms}, 
  year={2021},
  volume={2},
  number={},
  pages={1980-1995},
  doi={10.1109/OJCOMS.2021.3105933}}

@ARTICLE{AHO-14,
  author={Gao, Ning and Jin, Shi and Li, Xiao and Matthaiou, Michail},
  journal={IEEE Wireless Commun. Lett.}, 
  title={Aerial {RIS}-Assisted High Altitude Platform Communications}, 
  year={2021},
  volume={10},
  number={10},
  pages={2096-2100},
  doi={10.1109/LWC.2021.3091164}}

@ARTICLE{AHO-15,
  author={Shaik, Parvez and Kishore Garg, Kamal and Kumar Singya, Praveen and Bhatia, Vimal and Krejcar, Ondrej and Alouini, Mohamed-Slim},
  journal={IEEE Open J. Commun. Soc.}, 
  title={On Performance of Integrated Satellite {HAPS} Ground Communication: Aerial {IRS} Node vs Terrestrial {IRS} Node}, 
  year={2024},
  volume={5},
  number={},
  pages={3775-3791},
  doi={10.1109/OJCOMS.2024.3411314}}

@ARTICLE{AHO-16,
  author={Dolas, Kshitija and Bhatnagar, M. R.},
  journal={IEEE Trans. Aerosp. Electron. Syst.}, 
  title={On Performance of {IRS}-Assisted Hybrid Satellite–Terrestrial Cooperative Communication}, 
  year={2023},
  volume={59},
  number={2},
  pages={2020-2028},
  doi={10.1109/TAES.2022.3200930}}

@ARTICLE{AHO-17,
  author={Dolas, Kshitija and Bhatnagar, Manav R.},
  journal={IEEE Trans. Aerosp. Electron. Syst.}, 
  title={Cooperative Dual Aerial {IRS}-Aided Satellite System: A Performance Analysis}, 
  year={2026},
  volume={62},
  number={},
  pages={5336-5351},
  doi={10.1109/TAES.2026.3660247}}

@ARTICLE{AHO-18,
  author={Benaya, Ahmed M. and AlHajri, Mohamed I. and Ismail, Mahmoud H. and Hassan, Mohamed S. and Landolsi, Taha},
  journal={IEEE Open J. Commun. Soc.}, 
  title={Outage Analysis of {THz}-Enabled {IRS}-Assisted Multi-User Space–Air—Ground Integrated Networks}, 
  year={2025},
  volume={6},
  number={},
  pages={2599-2615},
  doi={10.1109/OJCOMS.2025.3556887}}

@ARTICLE{AHO-19,
  author={Ali, Zain and Althunibat, Saud and Hasna, Mazen Omar and Qaraqe, Khalid},
  journal={IEEE Trans. Aerosp. Electron. Syst.}, 
  title={Secure Sum-Rate Optimization in {RIS}-Assisted {RSMA} {HAP} Systems With Untrusted Users}, 
  year={2026},
  volume={62},
  number={},
  pages={5167-5182},
  doi={10.1109/TAES.2026.3657736}}

@ARTICLE{TVT_Tracy_Widom,
  author={Yilmaz, Tayfun and Ilhan, Haci and Gorcin, Ali and Yanikomeroglu, Halim},
  journal={IEEE Transactions on Vehicular Technology}, 
  title={Performance Analysis of {RIS}-Assisted Large-Scale {MIMO} Systems with Eigenmode Beamforming under Hardware Imperfections}, 
  year={2026},
  volume={},
  number={},
  pages={1-16},
  doi={10.1109/TVT.2026.3713803}}

\vfill
\end{document}